\documentclass[english,aps,prb,showpacs,superscriptaddress,floats,amsmath,amssymb,floatfix,nobalancelastpage,twocolumn,groupedaddress]{revtex4-2}
\usepackage[T1]{fontenc}
\usepackage{textcomp}
\usepackage[latin9]{inputenc}
\usepackage{color}
\usepackage{units}
\usepackage{amsmath}
\usepackage{amssymb}
\usepackage{graphicx}

\makeatletter

\usepackage{accents}
\allowdisplaybreaks[4]
\usepackage{orcidlink}

\makeatother

\usepackage{babel}
\graphicspath{{./}}
\begin{document}
\title{Magnetoelectric Raman Force on Shear Phonons in\\
a Frustrated van der Waals Bilayer Magnet}
\author{Wolfram Brenig \orcidlink{0000-0002-2968-6363}}
\email{w.brenig@tu-braunschweig.de}

\affiliation{Institute for Theoretical Physics, Technical University Braunschweig,
D-38106 Braunschweig, Germany}
\begin{abstract}
We show that the concept of coherent phonon generation by second order
response to incident electric laser fields, which is a hallmark of
pump-probe spectroscopy on conventional solids, can be expanded to
include frustrated quantum magnets. For that purpose, we analyze the
Raman force on the shear phonons of a frustrated magnetoelectric bilayer
spin system. The bilayer is a stacked triangular magnet, motivated
by recently emerging type-II van der Waals multiferroic transition
metal dihalides and comprises a spin system which allows for incommensurate
spiral order. The magnon excitations are treated by linear spin wave
theory. In the spiral state, a finite electric polarization is obtained
from the spin-current interaction which induces a coupling of the
magnons to the electric field. Scattering of the bilayer shear phonons
from the magnons is derived from a magnetoelastic energy. In this
scenario, a mixed three-point response function for the Raman force
is evaluated. We find it to be strongly anisotropic and very sensitive
to the magnon lifetime.
\end{abstract}
\maketitle

\section{Introduction\label{sec:Introduction}}

In this work, we connect two seemingly unrelated topics, namely the
physics of laser-pulse induced coherent phonons in nonmagnetic solids
with that of frustrated spin systems in layered type-II multiferroics.

Since the advent of femto- and picosecond lasers, pump-probe spectroscopy
(PPS) \citep{Weiner2009} has evolved into an important branch of
optics, used to investigate elementary excitations, equilibrium and
nonequilibrium properties of matter \citep{Cavalleri2001,Chollet2005,Fushitani2008,Saiki2017,Katsufuji2023}.
Coherent optical phonon (CP) excitations, which lead to oscillatory
reflectivity (transmissivity) changes, observable in the probe following
the pump pulse, are a prominent feature of PPS. They have been reported,
e.g., in insulating \citep{Cheng1991,Liu1995}, semiconducting \citep{Cheng1991,Cho1990,Pfeifer1992,Yamamoto1994,Dekorsy1995},
and correlated electron materials \citep{Chwalek1990,Mazin1994,Misochko2004}
(reviews in \citep{Kuett1992,Merlin1997}). For a CP to occur, its
harmonic oscillator equation, i.e., $\partial_{t}^{2}u(t){+}\allowbreak\Omega^{2}u(t){=}\allowbreak F(t)/\rho$,
with displacement coordinate $u$ and ionic mass density $\rho$,
requires a force $F(t)$ driving the phonon beyond a one-boson eigenstate.
The majority of CP in PPS can be understood by $F(t)$ resulting from
stimulated Raman scattering (SRS). There, briefly neglecting any time
dependence, the electric field pump of amplitude $E^{\alpha}$ implies
a potential energy of $V{\sim}\allowbreak-\chi^{\alpha\beta}\allowbreak E^{\alpha}E^{\beta}/2$
with a dielectric susceptibility $\chi^{\alpha\beta}$. This leads
to a force $F{=}\allowbreak\partial\chi^{\alpha\beta}/\partial u\,\allowbreak E^{\alpha}E^{\beta}/2$
on the phonon \citep{Cheng1991,Zeiger1992,Scholz1993,Kuznetsov1994,Garrett1996,Stevens2002},
dubbed \emph{Raman force} (RF). This force scales with the square
of the laser field. Depending on the time dependence of the pump-pulse,
the RF can be \emph{impulsive} $F(t){\sim}\delta(t)$, or \emph{displacive}
$F(t){\sim}\Theta(t)$. Impulsive excitations of CP (IECP) lead to
coherent oscillations around the phonon's original equilibrium position
($\sim u_{0}\sin(\Omega t)$), displacive excitations of CP (DECP)
oscillate around a shifted position ($\sim u_{0}(1-\cos(\Omega t))$).
DECP has been proposed to excite solely Raman active modes \citep{Zeiger1992}.

DECP has come under scrutiny recently in van der Waals (vdW) materials,
including bilayer graphene and transition-metal dichalcogenides, e.g.,
WTe$_{2}$ and MoTe$_{2}$. This is because DECP of shear phonons
implies a lateral shifting of the layers. For sufficiently large amplitude,
i.e. laser power, this may dynamically change the stacking order and
correspondingly the electronic properties \citep{Ho2006,Koshino2009}
which may also be of interest for Moir\'e variants of these materials.
In fact, DECP of shear phonons in MoTe$_{2}$ allows to optically
switch between the 1$T^{\prime}$-symmetric, normal semimetallic and
the $T_{d}$-symmetric, Type-II Weyl semimetal phases \citep{Zhang2019,Fukuda2020}.
A shear phonon induced change of stacking order has also been reported
for WTe$_{2}$ \citep{Ji2021}. This has revived theoretical interest
in the RF regarding shear phonons in bilayer graphene \citep{Rostami2022}.

As of today, standard approaches to the RF \citep{Cheng1991,Zeiger1992,Scholz1993,Kuznetsov1994,Garrett1996,Stevens2002}
are based on polarizing the charge degrees of freedom of an electronic
system by the laser. One main aim of this work is to go beyond this
limitation by considering the RF in magnetic systems modeled by quantum
spin systems. To couple the laser to the spin system we will consider
multiferroic quantum magnets which display the magnetoelectric effect
\citep{Fiebig2005,Tokura2014,Dong2015,Dong2019}. A second main motivation
of our work is the recent focus on transition metal dihalides (TMD)
which emerge as a new class of vdW materials, displaying noncollinear
magnetism and type-II multiferroicity down to the ultrathin limit
\citep{Botana2019,Zhang2020,Amoroso2020,Liu2020,Ju2021,Bikal2021,Song2022,Sodequist2023,Lebedev2023,Jiang2023,Gao2024,Song2025}.
The triangular vdW TMD NiI$_{2}$ is under particular scrutiny. Ab-inito
studies suggest this system to display layer-number and stacking dependent
intralayer incommensurate spiral spin orders \citep{NLiu2024} with
interlayer antiferromagnetism. This has been attributed to various
spin interactions \citep{Amoroso2020,NLiu2024,Bellaiche2023,Bennett2024},
but most directly can be captured by frustration from intralayer (anti)ferromagnetic
(third)first-nearest-neighbor exchange and antiferromagnetic interlayer
coupling \citep{Antao2024}. Due to the spiral order, NiI$_{2}$ allows
for spin-current - or Katsura Nagaosa Balatsky (KNB) - coupling to
electric fields \citep{Tokura2014,Katsura2005}. I.e., it displays
finite in-plane electric polarization \citep{Amini2024} and is a
type-II multiferroic \citep{NLiu2024,Antao2024} which is exactly
fit for our study. Finally, the third motivation of our work is that
similar to layered graphene and transition-metal dichalcogenides,
NiI$_{2}$ displays low-energy, Raman active shear modes, varying
with the number of layers \citep{Wu2023,Wu2024}.

Summarizing all of the preceding, we will analyze a magnetoelectric
RF on the shear phonons of a frustrated vdW bilayer magnet. For that,
we first describe the vdW spin model to be studied in Sec. \ref{sec:Spin-Model}
and detail its excitations in Sec. \ref{sec:Linear-spin-wave}. Then,
we discuss its magnetoelectric properties in Sec. \ref{sec:Polarization}.
The phonon modes of the vdW lattice are considered in Sec. \ref{sec:Shear-Phonons}.
The magnetoelastic coupling of the relevant phonon modes to the spin
system are modeled in Sec. \ref{sec:Magnon-shear-Phonon-Coupling}.
Finally, we combine all ingredients to evaluate the RF in Sec. \ref{sec:Magnetoelectric-Raman-Force}.
In Sec. \ref{sec:Conclusion}, we briefly conclude.

\begin{figure}
\begin{centering}
\includegraphics[width=0.5\columnwidth]{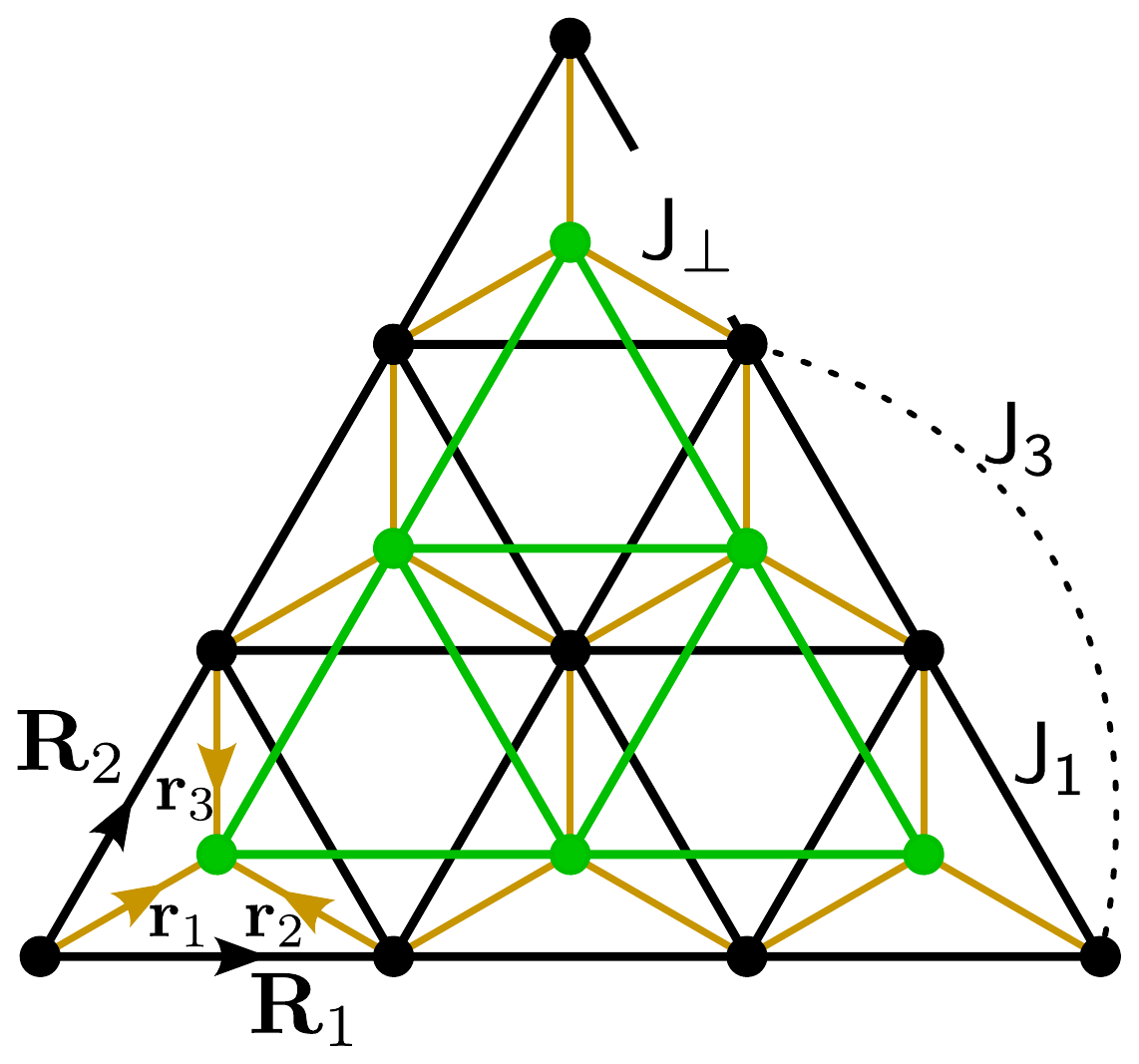}
\par\end{centering}
\caption{Top view of the triangular bilayer: black(green) bonds refer to bottom(top)
layer. Layers are vertically separated by $d$. Each layer has NN
and 3NN exchange, $J_{1}<0$ and $J_{3}>0$, respectively. Beige:
interlayer bonds with NN exchange $J_{\perp}>0$.\label{fig:model}}
\end{figure}

\section{Spin Model\label{sec:Spin-Model}}

Before starting, we emphasize that our study is not a material specific
analysis of NiI$_{2}$. Rather, we consider a model which may capture
genuine features of such triangular vdW magnets, however, we abstract
away from open issues regarding details of the ab-initio description,
exchange mechanisms, parameters, and stacking types \citep{Botana2019,Amoroso2020,Liu2020,Song2022,Sodequist2023,Gao2024,Song2025,NLiu2024,Bellaiche2023,Bennett2024}.
The main concepts of this work should be insensitive to details of
a particular model.

The spin system we consider is a bilayer of triangular lattices, stacked
such that the sites of its projection onto the $x,y$-plane form a
honeycomb lattice, dubbed AB-stacking, see Fig. \ref{fig:model}.
A fully aligned AA-stacking is not of interest to this study, since
the linear magnetoelastic coupling to shear phonons from a nearest-neighbor
interlayer exchange would vanish in this geometry, see Sec. \ref{sec:Magnon-shear-Phonon-Coupling}.
We use unit vectors $\boldsymbol{R}_{1}=(1,0)$, $\boldsymbol{R}_{2}=(1/2,\sqrt{3}/2)$
with a lattice constant $a\equiv1$ for the lower layer and the triangular
lattice sites are $\boldsymbol{R}_{\boldsymbol{l}1}=l_{1}\boldsymbol{R}_{1}+l_{2}\boldsymbol{R}_{2}$
with integer vector $\boldsymbol{l}=(l_{1},l_{2})$. The three vectors
directed towards the honeycomb basis sites around $\boldsymbol{R}_{\boldsymbol{0}1}$
are $\boldsymbol{r}_{1}=(1/2,1/(2\sqrt{3}))$, $\boldsymbol{r}_{2}=(-1/2,1/(2\sqrt{3}))$,
and $\boldsymbol{r}_{3}=(0,-1/\sqrt{3})$ and the location of the
sites of the upper layer projected onto the (x,y)-plane are $\boldsymbol{R}_{\boldsymbol{l}2}=\boldsymbol{R}_{\boldsymbol{l}1}+\boldsymbol{r}_{1}$.
The layers are separated by a distance $\boldsymbol{d}=(0,0,d)$,
i.e., the sites in the upper layer are at $\boldsymbol{R}_{\boldsymbol{l}2}+\boldsymbol{d}$.
For the treatment of the magnetism and the excitations, the layer
separation is irrelevant and we remain with a 2D hamiltonian
\begin{align}
H= & -\!\!\!\sum_{\langle\boldsymbol{l}i,\boldsymbol{m}i\rangle}\!\boldsymbol{S}_{\boldsymbol{l}i}{\cdot}\boldsymbol{S}_{\boldsymbol{m}i}+j\!\!\!\!\!\!\sum_{\langle\langle\langle\boldsymbol{l}i,\boldsymbol{m}i\rangle\rangle\rangle}\!\!\!\!\boldsymbol{S}_{\boldsymbol{l}i}{\cdot}\boldsymbol{S}_{\boldsymbol{m}i}+\nonumber \\
 & j_{\perp}\!\!\!\sum_{\langle\boldsymbol{l}2,\boldsymbol{m}1\rangle}\boldsymbol{S}_{\boldsymbol{l}2}{\cdot}\boldsymbol{S}_{\boldsymbol{m}1}\,.\label{mh}
\end{align}
We set all exchange energies as normalized to the \emph{absolute}
value of the nearest neighbor (NN) intralayer exchange $J_{1}$ which
is ferromagnetic (FM) in NiI$_{2}$. $j=J_{3}/|J_{1}|$ refers to
the third-nearest neighbor (3NN) intralayer, and $j_{\perp}=J_{\perp}/|J_{1}|$
to the NN interlayer exchange, both of which are antiferromagnetic
(AFM). The sums in the first two terms run over $i=1,2$, referring
to the lower and upper layer, respectively, while the third sum links
the layers. In accordance with NiI$_{2}$ we assume $j,j_{\perp}$
to be subdominant, i.e., $0\leq j,j_{\perp}\lesssim1$.

\begin{figure}[tb]
\centering{}\includegraphics[width=0.7\columnwidth]{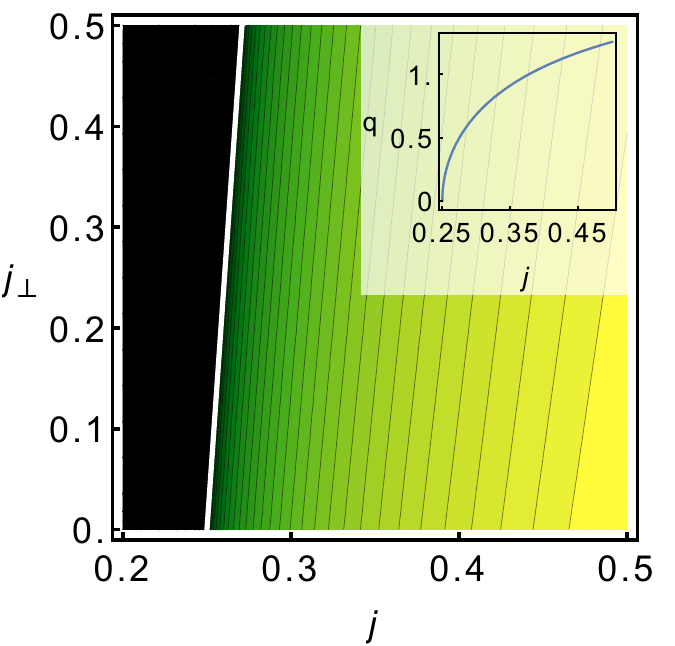}\caption{Contour plot of the pitch angle, tantamount to the classical phases
of the bilayer. FM (dark green), critical line (white), and ICS (green
to yellow gradient). Inset: Representative cut, displaying the pitch
at $j_{\perp}=0$. \label{fig:classph}}
\end{figure}

The intralayer classical ground state of the bilayer at $j_{\perp}=0$
is an FM for $j$ below a critical value of $j_{c}$. Above the latter,
and because of the frustrating action of $j$, it turns into an incommensurate
spiral (ICS). Owing to the triangular lattice symmetry, the direction
of the ICS pitch angle $\boldsymbol{q}$ is six-fold degenerate, separated
by an angle of $60{^\circ}$. Hereafter, we fix $\boldsymbol{q}$
to be one of these directions, namely $\boldsymbol{q}=(q{>}0,0)$.
Finite $j_{\perp}$ interlocks the ICS of the two layers, such that
the spins on the upper layer are almost completely antialigned to
the lower layer, including, however, the additional relative rotation
from the constant shift of $\boldsymbol{r}_{1}$ between the layers.
At $q=0$ this reduces to perfect antialignement. The classical energy
for spin $S$ per unit cell for the ICS configuration is $S^{2}E_{cl}=S^{2}((4j-\allowbreak2)\cos(q)+\allowbreak2j\cos(2q)-\allowbreak2(j_{\perp}+\allowbreak2)\cos(q/2)-\allowbreak j_{\perp})$.
The spiral pitch at the minimum of $E_{cl}$ is obtained from the
zeros of a cubic equation for $\cos(q)$. We find
\begin{alignat}{1}
 & q=\mathrm{Re}(\mathrm{acos}(\frac{4j(1-4j)+w_{+}+w_{-}}{24j^{2}}))\label{q}\\
 & w_{\pm}=j(a\pm b)^{1/3}\nonumber \\
 & a=54j(j_{\perp}{+}2)^{2}{-}(4j{+}2)^{3}\nonumber \\
 & b=6\sqrt{3}(j_{\perp}{+}2)({-}j(8{+}j(32j(2j{+}3){-}\nonumber \\
 & \hphantom{{aaaaaa\}}}3(3j_{\perp}{+}2)(3j_{\perp}{+}10))))^{1/2}\,,\nonumber 
\end{alignat}
where the cut of $z^{x}$ with fixed $x>0$ is on the negative real
$z$-axis. From this equation, the classical critical line for the
transition into the ICS can be extracted as $j_{\perp}=6(4j-1)$.
For $j_{\perp}=0$, the pitch simplifies to
\begin{equation}
q=2\,\mathrm{Re}[\mathrm{atan2}(\sqrt{2\frac{7j{-}\sqrt{j(j{+}2)}{-}1}{j}},\frac{j{+}\sqrt{j(j{+}2)}}{j})]\label{qs}
\end{equation}
The classical phase diagram is depicted in Fig. \ref{fig:classph}.
As to be expected, finite $j_{\perp}>0$ suppresses the tendency to
form an ICS. However, in the tricoordinated geometry of the AB stacking,
the effect of $j_{\perp}$ on supporting intralayer FM order in favor
of the ICS is weak only, and moreover much weaker than in a similar
model with AA-stacking \citep{Antao2024}, featuring triangular layers
directly on top of each other and with only a single perpendicular
interlayer exchange per unit cell.

\section{Linear spin wave theory\label{sec:Linear-spin-wave}}

Now we detail the magnon excitations of the bilayer. To begin we rotate
the ICS onto a locally FM coordinate frame with spins $\tilde{\boldsymbol{S}}_{\boldsymbol{l}i}$
and the classical order parameter $S$ pointing into the $z$-direction
\begin{equation}
\boldsymbol{S}_{\boldsymbol{l}i}{=}\left[\begin{array}{ccc}
0 & {\pm}\sin(\boldsymbol{q}{\cdot}\boldsymbol{R}_{\boldsymbol{l}i}) & {\mp}\cos(\boldsymbol{q}{\cdot}\boldsymbol{R}_{\boldsymbol{l}i})\\
0 & {\mp}\cos(\boldsymbol{q}{\cdot}\boldsymbol{R}_{\boldsymbol{l}i}) & {\mp}\sin(\boldsymbol{q}{\cdot}\boldsymbol{R}_{\boldsymbol{l}i})\\
-1 & 0 & 0
\end{array}\right]\tilde{\boldsymbol{S}}_{\boldsymbol{l}i}\,,\label{eq:RS}
\end{equation}
where the lower(upper) sign is for the lower(upper) layer, $i=1$($2$).
Thus, for the lower and upper layer the classical spiral is $S\,[\cos((l_{1}+l_{2}/2)q),\sin((l_{1}+l_{2}/2)q),0]$
and $S\,[{-}\cos((l_{1}+l_{2}/2+1/2)q),{-}\sin((l_{1}+l_{2}/2+1/2)q),0]$,
respectively. Describing quantum fluctuations off this state, we use
Holstein-Primakoff (HP) bosons. In terms of $\tilde{\boldsymbol{S}}_{\boldsymbol{l}i}$
they require only a\emph{ two site} magnetic unit cell, with
\begin{align}
\tilde{S}_{\boldsymbol{l}i}^{z}= & S-a_{\boldsymbol{l}i}^{\dagger}a_{\boldsymbol{l}i}^{\phantom{\dagger}}\,,\nonumber \\
\tilde{S}_{\boldsymbol{l}i}^{+}= & (2S-a_{\boldsymbol{l}i}^{\dagger}a_{\boldsymbol{l}i}^{\phantom{\dagger}})^{1/2}\,a_{\boldsymbol{l}i}^{\phantom{\dagger}}\,,\nonumber \\
\tilde{S}_{\boldsymbol{l}i}^{-}= & a_{\boldsymbol{l}i}^{\dagger}(2S-a_{\boldsymbol{l}i}^{\dagger}a_{\boldsymbol{l}i}^{\phantom{\dagger}})^{1/2}\,.\label{eq:HPB}
\end{align}
Performing the usual expansion of the Hamiltonian to leading $O(1/S)$
in terms of these HP boson, i.e., linear spin wave theory (LSWT),
we arrive at
\begin{equation}
H=\sum_{\boldsymbol{k}}\boldsymbol{A}_{\boldsymbol{k}}^{+}\boldsymbol{h}_{\boldsymbol{k}}\boldsymbol{A}_{\boldsymbol{k}}^{\phantom{+}}+NS(S+1)E_{cl}\,,\label{eq:H}
\end{equation}
where $N$ is the number of unit cells. $\boldsymbol{A}_{\boldsymbol{k}}^{+}=(a_{\boldsymbol{k}1}^{\dagger},\allowbreak a_{\boldsymbol{k}2}^{\dagger},\allowbreak a_{-\boldsymbol{k}1}^{\phantom{\dagger}},\allowbreak a_{-\boldsymbol{k}2}^{\phantom{\dagger}})$
is a boson spinor with $a_{\boldsymbol{k}n}^{\dagger}=\allowbreak\sum_{\boldsymbol{l}}\allowbreak\exp(i\boldsymbol{R}_{\boldsymbol{l}n}\cdot\boldsymbol{k})\allowbreak a_{\boldsymbol{l}n}^{\dagger}/\allowbreak N^{1/2}$
and
\begin{equation}
\boldsymbol{h}_{\boldsymbol{k}}=\left[\begin{array}{cccc}
\mathcal{A}_{\boldsymbol{k}} & \mathcal{D}_{\boldsymbol{k}} & \mathcal{B}_{\boldsymbol{k}} & \mathcal{E}_{\boldsymbol{k}}^{\star}\\
\mathcal{D}_{\boldsymbol{k}}^{\star} & \mathcal{A}_{\boldsymbol{k}} & \mathcal{E}_{\boldsymbol{k}} & \mathcal{B}_{\boldsymbol{k}}\\
\mathcal{B}_{\boldsymbol{k}} & \mathcal{E}_{\boldsymbol{k}}^{\star} & \mathcal{A}_{\boldsymbol{k}} & \mathcal{D}_{\boldsymbol{k}}\\
\mathcal{E}_{\boldsymbol{k}} & \mathcal{B}_{\boldsymbol{k}} & \mathcal{D}_{\boldsymbol{k}}^{\star} & \mathcal{A}_{\boldsymbol{k}}
\end{array}\right]
\end{equation}
with
\begin{align}
\mathcal{A}_{\boldsymbol{k}}= & \frac{1}{2}S((\cos(q){+}1)\cos(k_{x})(2j\cos(\sqrt{3}k_{y}){-}1){+}\nonumber \\
 & 2\cos(q)(j\cos(q)\cos(2k_{x}){-}2j{+}1){-}2j\cos(2q){-}\nonumber \\
 & 4\cos^{2}(\frac{q}{4})\cos(\frac{k_{x}}{2})\cos(\frac{\sqrt{3}k_{y}}{2}){+}4\cos(\frac{q}{2}){+}\nonumber \\
 & j_{\perp}(2\cos(\frac{q}{2}){+}1))+\nu\label{eq:hent1}\\
\mathcal{B}_{\boldsymbol{k}}= & S\sin^{2}(\frac{q}{4})(4\cos^{2}(\frac{q}{4})(\cos(k_{x})(2j((\cos(q){+}1)\times\nonumber \\
 & \cos(k_{x}){+}\cos(\sqrt{3}k_{y})){-}1){-}4j\cos^{2}(\frac{q}{2})\times\nonumber \\
 & \sin^{2}(k_{x})){-}2\cos(\frac{k_{x}}{2})\cos(\frac{\sqrt{3}k_{y}}{2}))\\
\mathcal{D}_{\boldsymbol{k}}= & Sj_{\perp}e^{\frac{ik_{y}}{2\sqrt{3}}}\sin^{2}(\frac{q}{4})\cos(\frac{k_{x}}{2})\\
\mathcal{E}_{\boldsymbol{k}}= & \frac{1}{2}Sj_{\perp}e^{\frac{ik_{y}}{\sqrt{3}}}(1{+}2e^{-\frac{i\sqrt{3}k_{y}}{2}}\cos^{2}(\frac{q}{4})\cos(\frac{k_{x}}{2}))\,.\label{eq:hent4}
\end{align}
Hereafter, bold faced spinor operators will also be referenced by
their components, using the notation $A_{\boldsymbol{k}\mu}^{(+)}$
with non-bold letters and subscripts $\mu=1,..4.$ The Hamiltonian
Eq. (\ref{eq:H}) can be diagonalized by means of a Bogoliubov transformation
$\boldsymbol{A}_{\boldsymbol{k}}^{+}=\boldsymbol{D}_{\boldsymbol{k}}^{+}\boldsymbol{U}_{k}^{+}$
onto diagonal bosons $\boldsymbol{D}_{\boldsymbol{k}}^{+}=(D_{\boldsymbol{k}1}^{+},..D_{\boldsymbol{k}4}^{+})=(d_{\boldsymbol{k}1}^{\dagger},\allowbreak d_{\boldsymbol{k}2}^{\dagger},\allowbreak d_{-\boldsymbol{k}1}^{\phantom{\dagger}},\allowbreak d_{-\boldsymbol{k}2}^{\phantom{\dagger}})$
for magnon with energies $\epsilon_{\boldsymbol{k}i=1,2}$. The Bogoliubov
transformation is paraunitary, i.e., $\boldsymbol{U}_{k}^{+}\boldsymbol{P}\boldsymbol{U}_{k}^{\phantom{+}}=\boldsymbol{P}$
with $\boldsymbol{P}$ diagonal and $\mathrm{diag}(\boldsymbol{P})=(1,1,-1,-1)$.
Moreover, $\boldsymbol{U}_{\boldsymbol{k}}^{+}\boldsymbol{h}_{\boldsymbol{k}}^{\phantom{+}}\boldsymbol{U}_{\boldsymbol{k}}^{\phantom{+}}$
is also diagonal with
\begin{equation}
\mathrm{diag}(\boldsymbol{U}_{\boldsymbol{k}}^{+}\boldsymbol{h}_{\boldsymbol{k}}^{\phantom{+}}\boldsymbol{U}_{\boldsymbol{k}}^{\phantom{+}})=\frac{1}{2}(\epsilon_{\boldsymbol{k}1},\epsilon_{\boldsymbol{k}2},\epsilon_{-\boldsymbol{k}1},\epsilon_{-\boldsymbol{k}2})\,.
\end{equation}
The magnon energies are the eigenvalues of the nonhermitian matrix
$\boldsymbol{P}\boldsymbol{h}_{\boldsymbol{k}}$ and can be expressed
analytically using Eqs. (\ref{eq:hent1}-\ref{eq:hent4})
\begin{align}
\epsilon_{\boldsymbol{k}i}= & 2[\mathcal{A}_{\boldsymbol{k}}^{2}{-}\mathcal{B}_{\boldsymbol{k}}^{2}{+}|\mathcal{D}_{\boldsymbol{k}}|^{2}{-}|\mathcal{E}_{\boldsymbol{k}}|^{2}{\mp}\nonumber \\
 & \hphantom{aaa}2[|\mathcal{F}_{\boldsymbol{k}}|^{2}{-}(\mathrm{Im}\mathcal{G}_{\boldsymbol{k}})^{2}]^{1/2}]^{1/2}\label{eq:ek}
\end{align}
with $\mathcal{F}_{\boldsymbol{k}}=\mathcal{A}_{\boldsymbol{k}}\mathcal{D}_{\boldsymbol{k}}-\mathcal{B}_{\boldsymbol{k}}\mathcal{E}_{\boldsymbol{k}}^{\star}$
and $\mathcal{G}_{\boldsymbol{k}}=\mathcal{D}_{\boldsymbol{k}}\mathcal{E}_{\boldsymbol{k}}$.
These energies are inversion and mirror symmetric, $\epsilon_{\boldsymbol{k}i=1,2}=\allowbreak\epsilon_{-\boldsymbol{k}i=1,2}$
and $\epsilon_{k_{x},k_{y}\,i=1,2}=\allowbreak\epsilon_{-k_{x},k_{y}\,i=1,2}=\allowbreak\epsilon_{k_{x},-k_{y}\,i=1,2},$respectively,
and therefore, in terms of the $\boldsymbol{D}_{\boldsymbol{k}}^{(+)}$
bosons
\begin{equation}
H=\sum_{\boldsymbol{k},i=1,2}\epsilon_{\boldsymbol{k}i}d_{\boldsymbol{k}i}^{\dagger}d_{\boldsymbol{k}i}^{\phantom{\dagger}}+E_{0}\,,\label{eq:digh}
\end{equation}
with\textcolor{red}{{} }$E_{0}=NS(S+1)E_{cl}+\sum_{\boldsymbol{k},i}\epsilon_{\boldsymbol{k}i}/2$,
where the second term is the zero-point energy of the harmonic Bose
gas encoded in Eq. (\ref{eq:digh}). Unfortunately, expressing the
Bogoliubov transformation in a manageable analytic form is not feasible
\citep{Bog1lay}. Therefore, in all later calculations, an explicit
use of $\boldsymbol{U}_{\boldsymbol{k}}$ is based on its numerical
evaluation, using Colpa's algorithm \citep{Colpa1978}. Related to
that, we caution that $\boldsymbol{U}_{\boldsymbol{k}}$ mixes bosons
at $\boldsymbol{k}$ and $-\boldsymbol{k}$. This has to be respected
by numerical procedures.

\begin{figure}[tb]
\begin{centering}
\includegraphics[width=0.48\columnwidth]{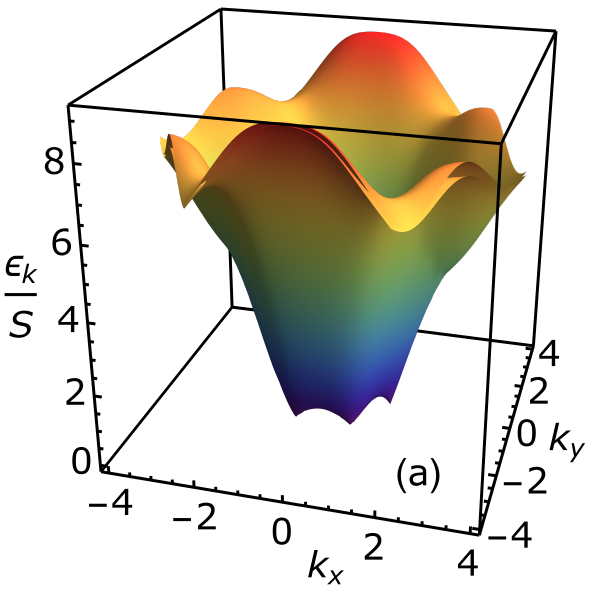}\includegraphics[width=0.48\columnwidth]{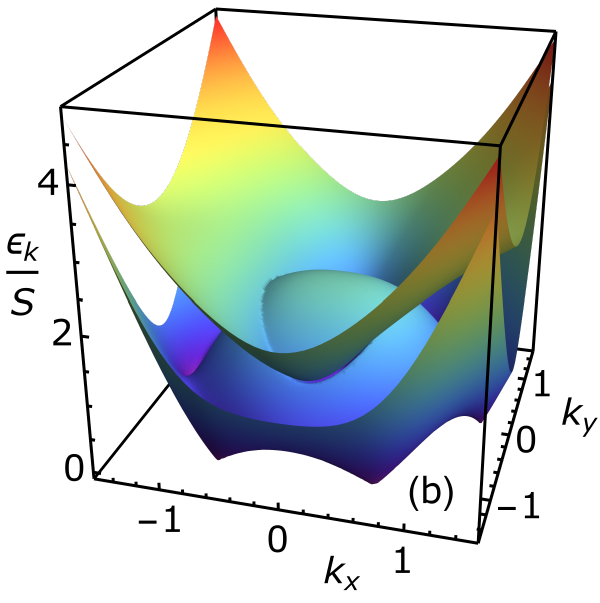}
\par\end{centering}
\centering{}\includegraphics[width=0.48\columnwidth]{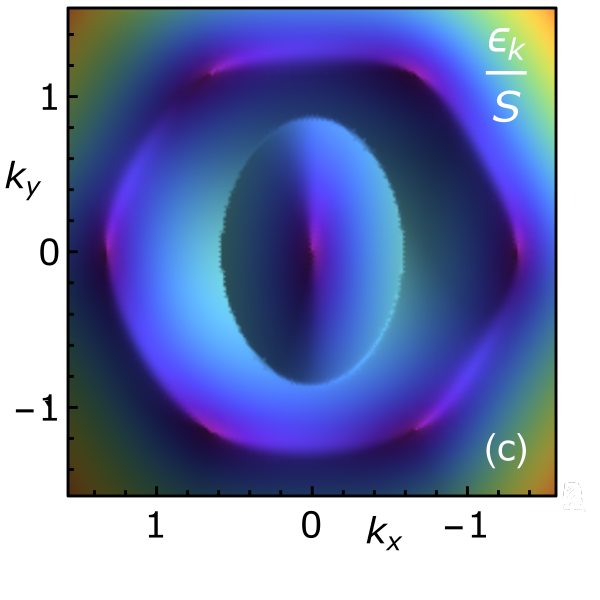}\includegraphics[width=0.48\columnwidth]{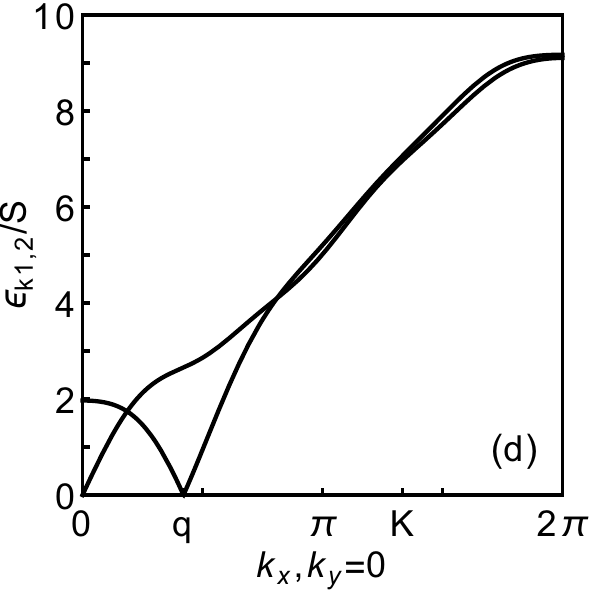}\caption{Magnon dispersion. (a) 3D plot in first BZ. (b) Blow up of (a) in
center of BZ. (c) Bottom-up view of (b). (d) Cut along straight line
through $K$-point. Parameters: $j=0.6$, $j_{\perp}=0.5$, and $\nu=0$.
For these $j$, $j_{\perp}$ the pitch is $q{\simeq}1.3217$\label{fig:magdis}}
\end{figure}

The last term $\nu$ in Eq. (\ref{eq:hent1}) requires some explanation.
It is a free parameter, added by hand. Its effect is to gap out the
zeros of the magnon dispersion $\propto\alpha\,\nu^{1/2}$, with $\alpha=O(1)$.
From the Mermin-Wagner-Hohenberg theorem, breaking a continuous symmetry
in $\ensuremath{D\leq2}$ and $\ensuremath{T\neq0}$, as in this work,
should not be possible. In practice, and for $D=2$, this tends to
manifest itself by weak finite-$T$ logarithmic singularities when
integrating over poles $\sim1/\epsilon_{\boldsymbol{k}i}^{2}$ arising
from Goldstone zeros. The cure for this has been documented in seminal
works on the square lattice Heisenberg antiferromagnet \citep{HH,CHN,Grempel,Tyc,Makivic,Hasenfratz00,Greven,Kastner98,Ronnow,Goff}.
These have shown that in $D=2$, finite-$T$ magnon propagation is
overdamped beyond a correlation length $\xi$. This introduces an
effective gap $\epsilon_{c}\sim v/\xi$ into the magnon dispersion,
where $v$ is a low-energy magnon velocity \citep{Takahashi,Auerbach1988,Barabanov1992}.
While analyzing such $\epsilon_{c}$ versus temperature and parameters
for our model is completely beyond the scope of this work, we will
use $\nu$ to phenomenologically introduce a gap $\epsilon_{c}\ll\mathrm{max}(\epsilon_{{\bf k}i})$
in some of the discussion. 

The dispersions of the two magnon branches are depicted in Fig. \ref{fig:magdis}
for a representative set of parameters. Several points should be noted.
First, and as to be expected, there is one acoustical and one optical
magnon branch, which can be seen best in panel (d) of the figure at
the $\Gamma$-point, $\boldsymbol{k}=\boldsymbol{0}$. Second, and
apart from the acoustical mode which is required by spin conservation,
there a six Goldstone modes due to the ICS, visible in all panels
but most evidently in panel (c), at the wave vectors $|\boldsymbol{q}_{i=1,..6}|=q$
with $\boldsymbol{q}_{1}=(q,0)$ and connected by $\pi/3$ rotations.
Finally, there are lines of degeneracy of the two magnon branches
in $\boldsymbol{k}$-space, e.g., clearly visible in panel (b). If
these degeneracies can be split by suitable additional interactions
and also if that may allow for topological magnon properties could
be of interest. For clarity we mention, that the straight line along
$k_{x}$ at $k_{y}=0$ in panel (c) does not pass through the $K$-point
perpendicular to a BZ boundary-line, i.e., the slope of $\epsilon_{\boldsymbol{k}i}$
can be nonzero there.

\section{Polarization\label{sec:Polarization}}

Type-II multiferroicity and finite electric polarization in bulk,
mono-, and multi-layered NiI$_{2}$ has been analyzed theoretically
as well as with experimental evidence provided by scanning tunneling
microscopy, circular dichroic Raman, birefringence, and second-harmonic-generation
measurements \citep{Ju2021,Song2022,Lebedev2023,Gao2024,Song2025,NLiu2024,Bennett2024,Antao2024,Amini2024}.
Here, we use the spin-current induced, or KNB coupling \citep{Tokura2014,Katsura2005}
to model type-II multiferroicity and the coupling to external dynamic
electric fields $\boldsymbol{E}(t)$. For the bilayer of our study,
this reads
\begin{align}
\lefteqn{H_{\mathrm{KNB}}(t)=-\boldsymbol{P}\cdot\boldsymbol{E}(t)}\label{eq:hknp}\\
 & \boldsymbol{P}/g=\sum_{\langle\boldsymbol{l}i,\boldsymbol{m}i\rangle}\!\boldsymbol{R}_{\boldsymbol{l}i,\boldsymbol{m}i}{\times}\boldsymbol{S}_{\boldsymbol{l}i}{\times}\boldsymbol{S}_{\boldsymbol{m}i}+\gamma\!\!\!\!\!\!\sum_{\langle\langle\langle\boldsymbol{l}i,\boldsymbol{m}i\rangle\rangle\rangle}\!\!\!\!\!\!\!\boldsymbol{R}_{\boldsymbol{l}i,\boldsymbol{m}i}{\times}\nonumber \\
 & \hphantom{aaaaaa}\boldsymbol{S}_{\boldsymbol{l}i}{\times}\boldsymbol{S}_{\boldsymbol{m}i}+\gamma_{\perp}\!\!\!\sum_{\langle\boldsymbol{l}2,\boldsymbol{m}1\rangle}\!\!\!\!\boldsymbol{R}_{\boldsymbol{l}2,\boldsymbol{m}1}{\times}\boldsymbol{S}_{\boldsymbol{l}2}{\times}\boldsymbol{S}_{\boldsymbol{m}1}\,,\label{eq:pknb}
\end{align}
where $\boldsymbol{P}$ is the electric polarization, $g$, $\gamma$
and $\gamma_{\perp}$ are effective NN, 3NN, and interlayer NN coupling
constants. For the remainder of this work, and similar to the unit
of energy, we normalize $\boldsymbol{P}$ to $g\equiv1$ in order
to abbreviate the notation. $\boldsymbol{R}_{\boldsymbol{l}i,\boldsymbol{m}j}=\boldsymbol{R}_{\boldsymbol{l}i}-\boldsymbol{R}_{\boldsymbol{m}j}$
are intersite separation vectors.

The classical KNB polarization $\boldsymbol{P}_{cl}$ results from
inserting the classical ICS into Eq. (\ref{eq:pknb}) using Eq. (\ref{eq:RS}).
This yields
\begin{equation}
P_{cl}^{y}{=}S^{2}((\gamma_{\perp}{-}2)\sin(\frac{q}{2}){-}2\sin(q)(2\gamma{+}4\gamma\cos(q){+}1))\label{eq:Pcl}
\end{equation}
with $P_{cl}^{x}=P_{cl}^{z}=0$. I.e., the classical polarization
is perpendicular to the ICS pitch vector and lies in the lattice plane
of the layers.

\begin{figure}[tb]
\centering{}\includegraphics[width=0.49\columnwidth]{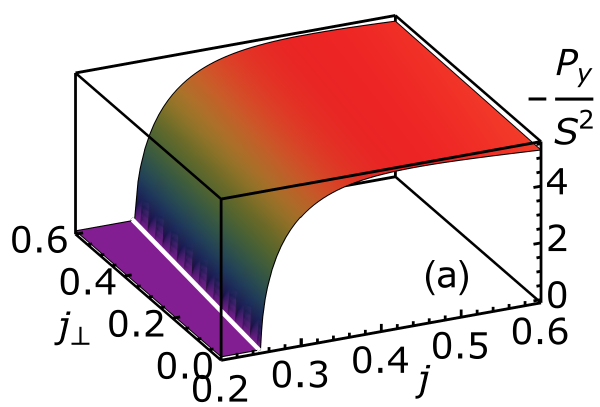}\includegraphics[width=0.49\columnwidth]{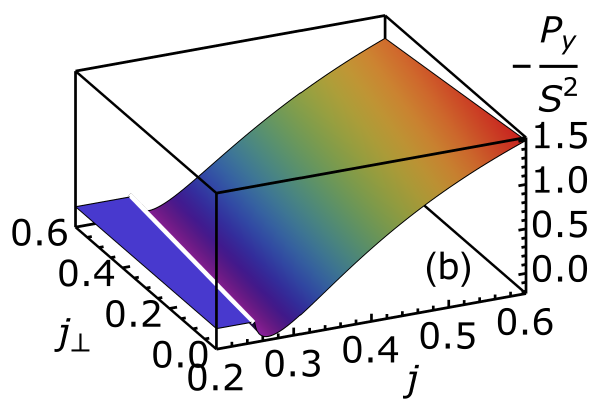}\caption{Representative plots of the classical polarization versus exchange
for (a) $\gamma=0.4$, $\gamma_{\perp}=0.2$ and (b) $\gamma=-0.3$,
$\gamma_{\perp}=0.2$. White line: classical critical line, as in
Fig. (\ref{fig:classph}).\label{fig:classP}}
\end{figure}

We emphasize that irrespective of the ongoing discussion about an
ab-initio description of vdW multiferroics like NiI$_{2}$ \citep{Botana2019,Amoroso2020,Liu2020,Song2022,Sodequist2023,Gao2024,Song2025,NLiu2024,Bellaiche2023,Bennett2024},
material specific values for $\gamma_{(\perp)}$ in units of $g$
are an open issue. This also pertains to the relative sign of these
parameters. A reasonable assumption is that the NN interlayer coupling
$g$ is largest, i.e., that $|\gamma_{(\perp)}|<1$. We adhere to
that. With this and because of the sole dependence via $(\gamma_{\perp}{-}2)$
in Eq. (\ref{eq:Pcl}), the influence of $\gamma_{\perp}$ is rather
weak. This is different for $\gamma$. Because $2\gamma{+}4\gamma\cos(q_{c})=6\gamma$,
a value of $-1\ll\gamma<-1/6$ can even lead to a sign reversal of
$P_{cl}^{y}$ in the vicinity of the critical line. Fig. \ref{fig:classP}
shows the classical polarization versus the exchange couplings for
two representative pairs of $\gamma_{(\perp)}$. The onset at the
critical line and also the sign reversal for suitable negative values
of $\gamma$ are clearly visible. Quantitative details of the Raman
force which we evaluate later will certainly vary with $\gamma_{(\perp)}$.
However, the main qualitative results will not. Therefore we refrain
from parameter studies and will use only a single fixed set of $0<\gamma_{(\perp)}<1$.

Next we consider the leading order quantum corrections to Eq. (\ref{eq:Pcl})
by inserting the HP bosons. To ease the calculations, we resort to
two simplifications. Summarizing them first, they comprise neglecting
(i) photon\--one\--magnon absortion and (ii) Umklapp scattering.
In more detail and for (i), to lowest order in $1/S$ a polarization
of type Eq. (\ref{eq:pknb}) certainly allows for linear mixing between
the dynamic electric field and single magnon excitations, i.e., so
called electromagnons. For all purposes of our analysis the wave vector
of the light $\boldsymbol{k}_{E}$ satisfies $\boldsymbol{k}_{E}=\boldsymbol{0}$.
Therefore, energy conservation implies that processes related to a
linear mixing have spectral weight only at the acoustical or optical
magnon energy at $\epsilon_{\boldsymbol{0}1}=0$ or $\epsilon_{\boldsymbol{0}2}$.
The former can be discarded since the external electric field has
a finite frequency. The latter will not be analyzed, since it is resonantly
confined to a single electric field frequency, i.e., $\epsilon_{\boldsymbol{0}2}$.
Rephrasing (i), we consider bilinear quantum corrections to $\boldsymbol{P}_{cl}$.

Turning to (ii), Umklapp scattering, the Hamiltonian being diagonal
in momentum space in the locally FM coordinate frame does not imply
this for other spinful operators. For them, momentum conservation
for the HP bosons may be broken by a finite number of integer multiples
of the ICS pitch vector $\boldsymbol{q}$. We want to avoid this complication.
\emph{A-posteriori,} we find (see Eqs. (\ref{eq:P2}-\ref{eq:p3}))
that the in-plane components of the bilinear quantum corrections to
$\boldsymbol{P}_{cl}$ from the triangular layers, i.e., from the
first two addends to $\boldsymbol{P}$ in Eq. (\ref{eq:pknb}), are
in fact momentum diagonal. Therefore, we consider only in-plane electric
fields $\boldsymbol{E}(t)\cdot\hat{\boldsymbol{z}}=0$. The bilinear
quantum correction to $\boldsymbol{P}_{cl}$ from $\gamma_{\perp}$,
however, is \emph{not} momentum diagonal. This leads to simplification
(ii): We consider only in-plane electric fields and neglect quantum
corrections from $\gamma_{\perp}$. We believe that none of these
simplifications will have a qualitative impact on our main findings.

With these remarks, we rotate Eq. (\ref{eq:pknb}) into the locally
FM frame, insert the HP bosons, and after some algebra find the bilinear
contribution in $\boldsymbol{A}_{\boldsymbol{k}}^{(+)}$, 
\begin{equation}
\mathcal{P}^{x(y)}=\sum_{\boldsymbol{k}}\boldsymbol{A}_{\boldsymbol{k}}^{+}\boldsymbol{p}_{\boldsymbol{k}}^{x(y)}\boldsymbol{A}_{\boldsymbol{k}}^{\phantom{+}}\,,\label{eq:P2}
\end{equation}
where
\begin{align}
\boldsymbol{p}_{\boldsymbol{k}}^{x}= & p_{\boldsymbol{k}1}\boldsymbol{L}_{1}\\
\boldsymbol{p}_{\boldsymbol{k}}^{y}= & p_{\boldsymbol{k}2}\boldsymbol{L}_{1}+p_{3}\boldsymbol{L}_{2}\,,
\end{align}
with
\begin{equation}
\boldsymbol{L}_{1}{=}\left[\begin{array}{cccc}
1 & 0 & -1 & 0\\
0 & 1 & 0 & -1\\
-1 & 0 & 1 & 0\\
0 & -1 & 0 & 1
\end{array}\right],\,\,\boldsymbol{L}_{2}{=}\left[\begin{array}{cccc}
0 & 0 & 1 & 0\\
0 & 0 & 0 & 1\\
1 & 0 & 0 & 0\\
0 & 1 & 0 & 0
\end{array}\right]\,,
\end{equation}
and
\begin{align}
p_{\boldsymbol{k}1}= & -S\frac{\sqrt{3}}{2}(\sin(\frac{k_{x}}{2})\sin(\frac{\sqrt{3}k_{y}}{2})\sin(\frac{q}{2}){+}\nonumber \\
 & \hphantom{aa}2\gamma\sin(k_{x})\sin(\sqrt{3}k_{y})\sin(q))\label{eq:p1}\\
p_{\boldsymbol{k}2}= & S\frac{1}{2}((2{-}\cos(\frac{k_{x}}{2})\cos(\frac{\sqrt{3}k_{y}}{2}))\sin(\frac{q}{2}){+}\nonumber \\
 & \hphantom{aa}(2{+}4\gamma{-}\cos(k_{x})(1{+}2\gamma\cos(\sqrt{3}k_{y})){+}\nonumber \\
 & \hphantom{aa}8\gamma\cos(q))\sin(q){-}2\gamma\cos(2k_{x})\sin(2q))\label{eq:p2}\\
p_{3}= & S((1{+}2\gamma{+}4\gamma\cos(q))\sin(q){+}\sin(\frac{q}{2}))\,.\label{eq:p3}
\end{align}
The polarization matrices $\boldsymbol{p}_{\boldsymbol{k}=k_{x},k_{y}}^{x(y)}$
satisfies the symmetries: $\boldsymbol{p}_{\,k_{x},k_{y}}^{x}{=}\allowbreak-\boldsymbol{p}_{\,-k_{x},k_{y}}^{x}$
and $\boldsymbol{p}_{\,k_{x},k_{y}}^{x}{=}\allowbreak-\boldsymbol{p}_{\,k_{x},-k_{y}}^{x}$,
as well as $\boldsymbol{p}_{\,k_{x},k_{y}}^{y}{=}\allowbreak\boldsymbol{p}_{\,-k_{x},k_{y}}^{y}$
and $\boldsymbol{p}_{\,k_{x},k_{y}}^{y}{=}\allowbreak\boldsymbol{p}_{\,k_{x},-k_{y}}^{y}$.
I.e., $\boldsymbol{p}_{\boldsymbol{k}}^{y}$($\boldsymbol{p}_{\boldsymbol{k}}^{x}$)
is (anti)symmetric under $x$ and $y$ mirroring.

For the remainder of this work, and to clarify the notation, we consider
an electric field with a fixed polarization into either $x$- or $y$-direction.
I.e., multiple superscripts of type $x(y)$ refer to either $x$ or
$y$ for all of them.

\section{Shear Phonons\label{sec:Shear-Phonons}}

In this section, we derive the planar vibrations of the bilayer including
its shear modes. As for the other parts of this work, we point out
that, while low-energy, Raman active shear modes have been observed
in NiI$_{2}$ which vary with the number of layers \citep{Wu2023,Wu2024},
our analysis of such modes for an AB-stacked bilayer does aim at a
one-to-one description of an experimental situation. The planar deformation
vectors are $\boldsymbol{\delta}=(\delta_{\boldsymbol{l}i}^{x},\delta_{\boldsymbol{l}i}^{y})$,
with $i=1,2$ for the lower and upper layer, respectively. We use
isotropic harmonic NN potentials between the triangular intralayer
sites, as well as between the NN interlayer sites, implying two force
constants $c_{1,2}$
\begin{align}
V= & \frac{c_{1}}{2}\sum_{\langle\boldsymbol{l}i,\boldsymbol{m}i\rangle}(|\boldsymbol{R}_{\boldsymbol{l}i}+\boldsymbol{\delta}_{\boldsymbol{l}i}-\boldsymbol{R}_{\boldsymbol{m}i}-\boldsymbol{\delta}_{\boldsymbol{m}i}|-1)^{2}+\nonumber \\
 & \frac{c_{2}}{2}\sum_{\langle\boldsymbol{l}2,\boldsymbol{m}1\rangle}(|\boldsymbol{R}_{\boldsymbol{l}2}+\boldsymbol{\delta}_{\boldsymbol{l}2}-\boldsymbol{R}_{\boldsymbol{m}1}-\boldsymbol{\delta}_{\boldsymbol{m}1}|-\frac{1}{\sqrt{3}})^{2}\,.\label{eq:1}
\end{align}
This setup is reminiscent of the description of planar phonons in
single layer graphene, assuming NN and NNN forces. On a technical
level, however, graphene tends to be considered with not only central-force
potentials \citep{Woods2000}. Such details are not relevant for the
present considerations. On a semantic level, for a single layer of
graphene the notion of a shear mode is meaningless.

\begin{figure}[tb]
\centering{}\includegraphics[width=0.47\columnwidth]{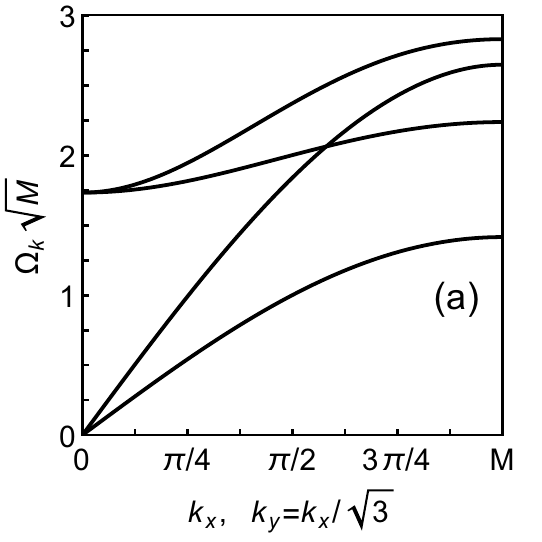}\hskip 0.cm\includegraphics[width=0.47\columnwidth]{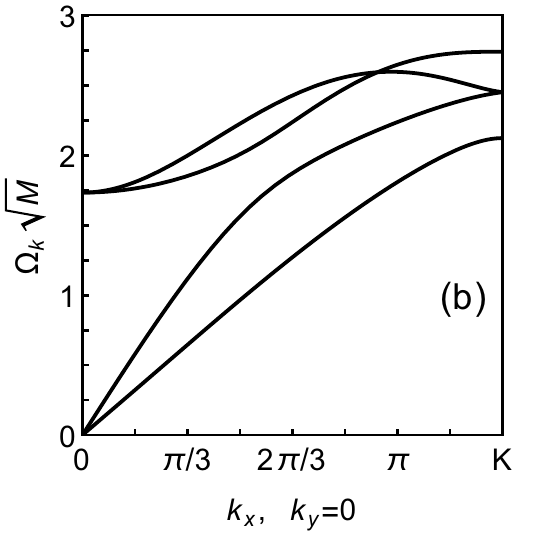}\caption{Planar phonon dispersions for $c_{1,2}=1$ and for two representative
directions in the BZ, (a) $\Gamma-M$, (b) $\Gamma-K$.\label{fig:phondisp}}
\end{figure}

In principle, and since the wave vector of the electric field is zero,
only the phonons at the BZ-center are needed for our purpose. Yet,
for completeness, we briefly derive the entire dispersions next. Expanding
Eq. (\ref{eq:1}) to $O(2)$ in the displacements and introducing
the Fourier transform $\boldsymbol{\delta}_{\boldsymbol{k}}^{\dagger}=\allowbreak\sum_{\boldsymbol{l}}(\exp(-i\boldsymbol{k}{\cdot}\allowbreak\boldsymbol{R}_{\boldsymbol{l}1}\allowbreak)\boldsymbol{\delta}_{\boldsymbol{l}1},\allowbreak\exp(-i\boldsymbol{k}{\cdot}\boldsymbol{R}_{\boldsymbol{l}2})\boldsymbol{\delta}_{\boldsymbol{l}2})$
we arrive at
\begin{equation}
V=\frac{1}{2}\sum_{\boldsymbol{k}}\boldsymbol{\delta}_{\boldsymbol{k}}^{\dagger}V(\boldsymbol{k})\boldsymbol{\delta}_{\boldsymbol{k}}\label{eq:2}
\end{equation}
with the dynamical matrix
\begin{align}
V(\boldsymbol{k})= & c_{1}\left[\begin{array}{cc}
v_{1}(\boldsymbol{k}) & \begin{array}{cc}
0 & 0\\
0 & 0
\end{array}\\
\begin{array}{cc}
0 & 0\\
0 & 0
\end{array} & v_{1}(\boldsymbol{k})
\end{array}\right]+c_{2}\left[\begin{array}{cc}
\begin{array}{cc}
\frac{3}{2} & 0\\
0 & \frac{3}{2}
\end{array} & v_{2}(\boldsymbol{k})\\
v_{2}(\boldsymbol{k})^{\dagger} & \begin{array}{cc}
\frac{3}{2} & 0\\
0 & \frac{3}{2}
\end{array}
\end{array}\right]\nonumber \\
v_{1,11}(\boldsymbol{k})= & 3{-}2\cos(k_{x}){-}\cos(k_{x}/2)\cos(\sqrt{3}k_{y}/2)\nonumber \\
v_{1,12}(\boldsymbol{k})= & v_{1,21}(\boldsymbol{k})=\sqrt{3}\sin(k_{x}/2)\sin(\sqrt{3}k_{y}/2)\nonumber \\
v_{1,22}(\boldsymbol{k})= & 3v_{1,11}(\boldsymbol{k}){+}6\cos(k_{x}){-}6\nonumber \\
v_{2,11}(\boldsymbol{k})= & {-}3\cos(k_{x}/2)\exp(ik_{y}/(2\sqrt{3}))/2\nonumber \\
v_{2,12}(\boldsymbol{k})= & v_{2,21}(\boldsymbol{k})=-i\sqrt{3}\sin(k_{x}/2)\exp(ik_{y}/(2\sqrt{3}))/2\nonumber \\
v_{2,22}(\boldsymbol{k})= & {-}\exp(-ik_{y}/\sqrt{3}){+}v_{2,11}(\boldsymbol{k})/3\,.\label{eq:3}
\end{align}
For $c_{2}=0$ or for $c_{1}=0$, i.e., for two decoupled triangular
or for the honeycomb lattice with both only NN force constants, the
phonon dispersions $\Omega_{\boldsymbol{k},\mu=1,..4}$ resulting
from Eqs. (\ref{eq:2}) and (\ref{eq:3}) can be obtained analytically
\begin{align}
\lefteqn{\frac{M\Omega_{\boldsymbol{k},\mu=1,..4}^{2}|_{c_{1}=0}}{c_{2}}{=}3,0,\{3{\pm}[3{+}2\cos(k_{x}){+}} & \hphantom{aaaaaaaaaaaaaaaaaaaaaaaaaaaaaaaaaaaaaaaa}\label{whon}\\
\lefteqn{\hphantom{aa}2\cos(\frac{k_{x}}{2}{+}\frac{\sqrt{3}k_{y}}{2}){+}2\cos(\frac{k_{x}}{2}{-}\frac{\sqrt{3}k_{y}}{2})]^{1/2}\}/2}\nonumber \\
\nonumber \\\lefteqn{\frac{M\Omega_{\boldsymbol{k},\mu=1,2}^{2}|_{c_{2}=0}}{c_{1}}=\frac{1}{2}(3{-}\cos(k_{x}){-}2\cos(\frac{k_{x}}{2})}\label{wtri}\\
\lefteqn{\hphantom{aa}\cos(\frac{\sqrt{3}k_{y}}{2}){\pm}\frac{1}{\sqrt{2}}[3{-}\cos(k_{x}){+}\cos(2k_{x}){-}4\cos(\frac{k_{x}}{2})\times}\nonumber \\
\lefteqn{\hphantom{aa}\cos(k_{x})\cos(\frac{\sqrt{3}k_{y}}{2}){+}(2\cos(k_{x}){-}1)\cos(\sqrt{3}k_{y})]^{1/2})\,,}\nonumber 
\end{align}
where $M$ is the effective ion mass and the longitudinal and transverse
modes of the decoupled triangular lattice layers, i.e., $\Omega_{\boldsymbol{k},\mu=1,2}|_{c_{2}=0}$
in Eq.(\ref{wtri}) are both twofold degenerate. For arbitrary $c_{1,2}$,
the phonon frequencies can be obtained numerically and are depicted
along two representative directions in Fig. \ref{fig:phondisp}.

\begin{figure}[tb]
\centering{}\includegraphics[width=0.7\columnwidth]{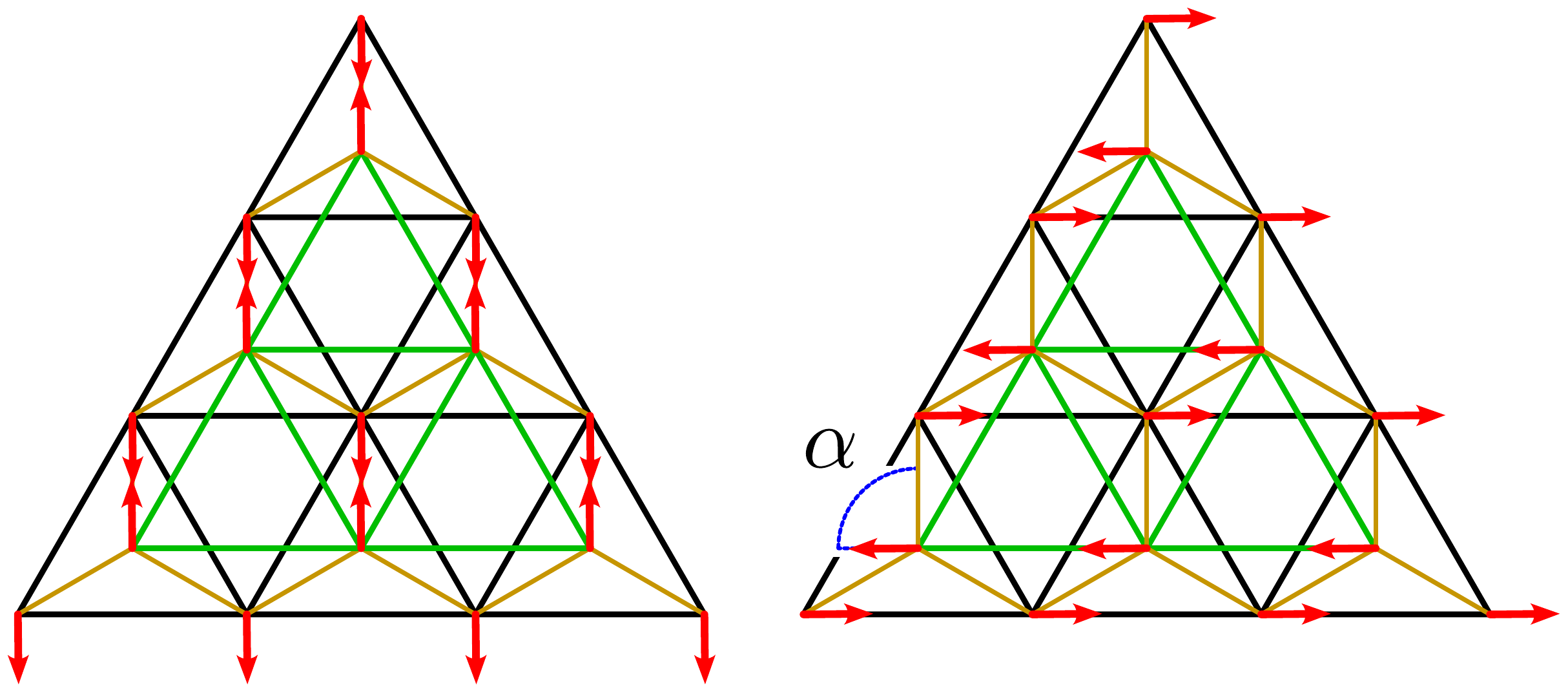}\caption{The two orthogonal homogeneous shear mode patterns. $\alpha$ denotes
the angle between the red amplitude vectors and the $y$-axis: transverse
$\alpha=0$, longitudinal $\alpha=\pi/2$.\label{fig:optmod}}
\end{figure}

For the remainder of this work, only the twofold degenerate optical
phonons at the $\Gamma$-point, $\boldsymbol{k}=\boldsymbol{0}$,
are relevant. They correspond to shear motions of the two layers with
respect to each other. The phonon frequencies and eigenvectors at
the $\Gamma$-point follow from
\begin{equation}
V(\boldsymbol{0})=V(\boldsymbol{0})|_{c_{1}=0}=\frac{3c_{2}}{2}\left[\begin{array}{cccc}
1 & 0 & -1 & 0\\
0 & 1 & 0 & -1\\
-1 & 0 & 1 & 0\\
0 & -1 & 0 & 1
\end{array}\right]\,,\label{opmod}
\end{equation}
where the first equality indicates that the intralayer motion is rigid
at $\boldsymbol{k}=\boldsymbol{0}$ and only the interlayer forces
contribute. Therefore the optical frequencies have already been given
in Eq. (\ref{whon}) and are $\sqrt{M}\Omega_{1,2}=\sqrt{3}c_{2}$.
An orthonormal set of phonon eigenvectors in the \emph{optical} subspace
derives from Eq. (\ref{opmod}) as $\boldsymbol{\delta}_{\boldsymbol{0},t}^{\dagger}=(0,-1,0,1)/\sqrt{2}$
and $\boldsymbol{\delta}_{\boldsymbol{0},l}^{\dagger}=(1,0,-1,0)/\sqrt{2}$.
We label them transverse (t) and longitudinal (l) hereafter. They
are depicted in Fig. \ref{fig:optmod}.

\section{Magnon shear-Phonon Coupling\label{sec:Magnon-shear-Phonon-Coupling}}

This section provides a description of the coupling between magnons
and the shear-phonons. We assume a bond-stretching type of magnetoelastic
energy
\begin{equation}
H_{sp}=\sum_{\langle\boldsymbol{l}2,\boldsymbol{m}1\rangle,\alpha=l,t}\!\!\!\!\!\!\!\tilde{j}_{\perp}\,\frac{\partial r_{\langle\boldsymbol{l}2,\boldsymbol{m}1\rangle}}{\partial u^{\alpha}}\,\boldsymbol{S}_{\boldsymbol{l}2}{\cdot}\boldsymbol{S}_{\boldsymbol{m}1}\,u^{\alpha}\,,
\end{equation}
where $\partial r_{\langle\boldsymbol{l}2,\boldsymbol{m}1\rangle}$
refers to the change of length of the planar NN-vectors $\boldsymbol{r}_{i{=}1,..3}$
connecting the basis sites of the planar honeycomb lattice projection
and 
\begin{equation}
\tilde{j}_{\perp}=\frac{1}{\sqrt{1+3d^{2}}}\,\left.\frac{\partial J_{\perp}}{\partial L}\right|_{L=\sqrt{d^{2}+1/3}}\,
\end{equation}
is the magnetoelastic coupling, scaled to include the interlayer distance
$d$, in units of the the lattice constant. $u^{\alpha}$ is the classical
amplitude of the optical phonon mode of polarization $\alpha$, shown
in Fig. \ref{fig:optmod}. The sum on $\alpha=l,t=\pi/2,0$ refers
to the angle of the two orthogonal optical modes with the $y$-axis,
see Fig. \ref{fig:optmod}. In principle, and by linear combination
of these degenerate modes, $\alpha$ can take on any value $\in[0,\pi[$.
For arbitrary $\alpha$, we abbreviate the projection of the phonon
amplitude onto the the basis vectors as
\begin{align}
\boldsymbol{g}^{\alpha}\equiv\frac{\partial r_{\langle\boldsymbol{l}2,\boldsymbol{0}1\rangle}}{\partial u^{\alpha}}=(g_{1}^{\alpha},g_{2}^{\alpha},g_{3}^{\alpha})=\hphantom{aaaaaaaaaaaaaaa}\\
(\cos(\alpha){-}\sqrt{3}\sin(\alpha),\cos(\alpha){+}\sqrt{3}\sin(\alpha),{-}2\cos(\alpha))\,,\nonumber 
\end{align}
where the three components $g_{i{=}1,2,3}^{\alpha}$ refer to the
three NN interlayer bonds $\langle\boldsymbol{l}2,\boldsymbol{0}1\rangle$.
Before evaluating the magnon-phonon coupling, we mention that, in
principle, a linear contribution to this is possible. This, however,
cannot lead to a nonlinear coupling to the electric field required
for the RF. Therefore, we expand to second order in the magnons and
consider only bilinear magnon-phonon coupling. We find
\begin{align}
H_{sp}= & \frac{1}{\sqrt{2M\Omega}}\sum_{\boldsymbol{k},\alpha=l,t}\boldsymbol{A}_{\boldsymbol{k}}^{+}\boldsymbol{\mathfrak{F}}_{\boldsymbol{k}}^{\alpha}\boldsymbol{A}_{\boldsymbol{k}}^{\phantom{+}}(b_{\alpha}^{\dagger}+b_{\alpha}^{\phantom{\dagger}})\label{eq:spcpl}\\
\boldsymbol{\mathfrak{F}}_{\boldsymbol{k}}^{\alpha}= & \tilde{j}_{\perp}\left[\begin{array}{cccc}
\mathfrak{f}_{\boldsymbol{k}1}^{\alpha} & \mathfrak{f}_{\boldsymbol{k}2}^{\alpha} & 0 & \mathfrak{f}_{\boldsymbol{k}3}^{\alpha\star}\\
\mathfrak{f}_{\boldsymbol{k}2}^{\alpha\star} & \mathfrak{f}_{\boldsymbol{k}1}^{\alpha} & \mathfrak{f}_{\boldsymbol{k}3}^{\alpha} & 0\\
0 & \mathfrak{f}_{\boldsymbol{k}3}^{\alpha\star} & \mathfrak{f}_{\boldsymbol{k}1}^{\alpha} & \mathfrak{f}_{\boldsymbol{k}2}^{\alpha}\\
\mathfrak{f}_{\boldsymbol{k}3}^{\alpha} & 0 & \mathfrak{f}_{\boldsymbol{k}2}^{\alpha\star} & \mathfrak{f}_{\boldsymbol{k}1}^{\alpha}
\end{array}\right]\nonumber 
\end{align}
with matrix elements
\begin{align}
\mathfrak{f}_{\boldsymbol{k}1}^{\alpha}= & \frac{1}{2}S((g_{1}^{\alpha}+g_{2}^{\alpha})\cos(\frac{q}{2})+g_{3}^{\alpha})\nonumber \\
\mathfrak{f}_{\boldsymbol{k}2}^{\alpha}= & \frac{1}{2}S(g_{2}^{\alpha}+g_{1}^{\alpha}e^{ik_{x}})e^{-\frac{i(3k_{x}-\sqrt{3}k_{y})}{6}}\sin^{2}(\frac{q}{4})\label{spc}\\
\mathfrak{f}_{\boldsymbol{k}3}^{\alpha}= & \frac{1}{2}S(\cos^{2}(\frac{q}{4})(g_{1}^{\alpha}{+}g_{2}^{\alpha}e^{ik_{x}})e^{{-}\frac{i(3k_{x}{+}\sqrt{3}k_{y})}{6}}{+}g_{3}^{\alpha}e^{\frac{ik_{y}}{\sqrt{3}}})\,.\nonumber 
\end{align}
Evidently, the dependence on $\alpha$ renders the coupling to the
shear-modes anisotropic. The expressions in Eq. (\ref{spc}) are valid
for arbitrary $\alpha$, however, we remain with $\alpha=0$ and $\pi/2$,
i.e., transverse and longitudinal. Here, the matrices $\boldsymbol{\mathfrak{F}}_{\boldsymbol{k}=k_{x},k_{y}}^{0(\pi/2)}$
satisfy the symmetries: $\boldsymbol{\mathfrak{F}}_{k_{x},k_{y}}^{0}{=}\allowbreak\boldsymbol{\mathfrak{F}}_{-k_{x},k_{y}}^{0}$
and $\boldsymbol{\mathfrak{F}}_{k_{x},k_{y}}^{0}{=}\allowbreak(\boldsymbol{\mathfrak{F}}_{k_{x},-k_{y}}^{0})^{T}$,
as well as $\boldsymbol{\mathfrak{F}}_{k_{x},k_{y}}^{\pi/2}{=}\allowbreak-\boldsymbol{\mathfrak{F}}_{-k_{x},k_{y}}^{\pi/2}$
and $\boldsymbol{\mathfrak{F}}_{k_{x},k_{y}}^{\pi/2}{=}\allowbreak-(\boldsymbol{\mathfrak{F}}_{k_{x},-k_{y}}^{\pi/2})^{T}$.
I.e., the coupling to the transverse(longitudinal) phonon is (anti)symmetric
under $x$ and $y$ mirroring, including a transposition for the $y$-direction.
We mention that the diagonal entries of $\boldsymbol{\mathfrak{F}}_{\boldsymbol{k}}^{\alpha}$
are very different for the two cases, namely, $\mathfrak{F}_{\boldsymbol{k}ii}^{0}=\allowbreak S(\cos(q/2)-1)$,
while $\mathfrak{F}_{\boldsymbol{k}ii}^{\pi/2}=0$.

\section{Magnetoelectric Raman Force\label{sec:Magnetoelectric-Raman-Force}}

The spin phonon coupling of Eq. (\ref{eq:spcpl}) can be written as
$H_{sp}=\sum_{\alpha}F^{\alpha}u^{\alpha}$, where the operator $F^{\alpha}$
can be viewed as a force acting on the phonon coordinate. The thermodynamic
average of this force is driven by the electric field via the spin-current
coupling Eq. (\ref{eq:hknp}). As motivated in Sec. (\ref{sec:Introduction})
for the DECP, one seeks for a response to \emph{dynamic} electric
(laser) fields which allows to produce DC forces, i.e., \emph{rectified}
forces which act on the phonon coordinates. These can shift the ionic
``equilibrium'' positions.

Response of the phonons can in principle start at $O(1)$ of the electric
field. However, this generates AC forces only at the incoming electric
field frequency. This is not of interest in the present context. In
terms of spectroscopy notions, such response could be called ``optical''.
At $O(2)$ of the electric field, parts of the incoming frequencies
can combine to generate a DC response. This is of interest for the
present study. Following the literature, the $O(2)$-effect is called
a ``Raman'' response and the forces are Raman forces.

\begin{figure}[tb]
\centering{}\includegraphics[width=0.4\columnwidth]{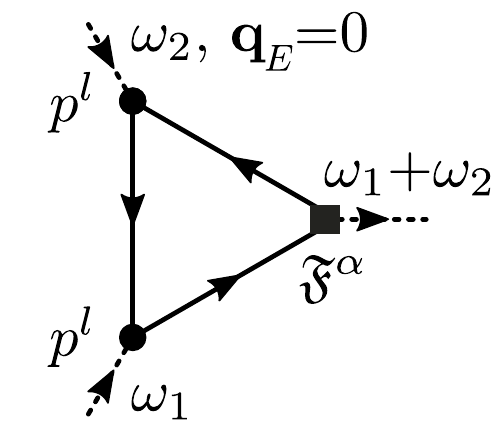}\caption{Diagram for the
  magnetoelectric Raman force response function
  $\chi^{\alpha l}_{2}(\omega_{1},\omega_{2})$ with $l=x(y)$ and $\alpha=l,t$.
  The solid lines carry an index $\mu=1,..4$, referring to $G_{1,..4}(\boldsymbol{k},i\omega_{n})$
of Eq. (\ref{G0}), the solid dots(square) refer to the symmetrized
Bogoliubov transforms of the $4\times4$ polarization(force) operator
matrices from Eq. (\ref{eq:P2})(Eq. (\ref{eq:spcpl})), respectively.
\label{fig:RFdiag}}
\end{figure}

We emphasize that we will consider the Raman force for arbitrary net
frequencies. It is obtained from the Fourier transform into the 2D
frequency plane of the $O(2)$ retarded nonlinear response function
$\tilde{\chi}_{2}^{\alpha l}(t,t_{1},t_{2})=i^{2}\Theta(t-t_{1})\Theta(t_{1}-t_{2})\allowbreak\langle[[F^{\alpha}(t),\allowbreak\mathcal{P}^{l}(t_{1})],\mathcal{P}^{l}(t_{2})]\rangle$
\citep{Butcher1990}, convoluted with the electric field. The $N$-fold
time integrations in the perturbative expansion of $\langle F^{\alpha}\rangle(t)$
at any order $N$ in $\mathcal{P}^{l}E^{l}(t)$, $l=x(y)$, are totally
symmetric with respect to any permutation of the $N$ time arguments
\citep{Butcher1990}, with $N=2$ in the present case. This is termed
intrinsic permutation symmetry. In turn, any $N$-th order contributions
to $\langle F^{\alpha}\rangle(t)$ requires only the fully symmetrized
response function $\chi_{N}^{\alpha l}(t,t_{1},\allowbreak\dots,t_{n})=\allowbreak\sum_{M}\tilde{\chi}_{N}(t,t_{M(1)},\allowbreak\dots,t_{M(N)})/N!$
to be evaluated, where $M$ labels all permutations. The Fourier transform
of $\chi_{N}^{\alpha l}(t,t_{1},\allowbreak\dots,t_{n})$ can be obtained
from analytic continuation to the real axis of the Matsubara frequency
transform of the fully connected contractions of the imaginary time
propagator $\chi_{n}^{\alpha l}(\tau_{n},\dots\tau_{1})=\langle T_{\tau}(\allowbreak F^{\alpha}(\tau_{n})\allowbreak\dots\mathcal{P}^{l}(\tau_{1})\mathcal{P}^{l})\rangle$
\citep{Evans1966,Rostami2021}.

The diagrammatic evaluation of $\chi_{2}$ is performed in the diagonal
HP bosons basis. This involves calculation of contractions from operator
groupings of type
\begin{align}
\sum_{\dots\boldsymbol{k}_{l}\dots\boldsymbol{k}_{m}\dots} & \langle T_{\tau}(\dots\boldsymbol{D}_{\boldsymbol{k}_{l}}^{+}(\tau_{l})\boldsymbol{s}_{\boldsymbol{k}_{l}}^{l}\boldsymbol{D}_{\boldsymbol{k}_{l}}(\tau_{l})\dots\nonumber \\
 & \dots\boldsymbol{D}_{\boldsymbol{k}_{m}}^{+}(\tau_{m})\boldsymbol{s}_{\boldsymbol{k}_{m}}^{m}\boldsymbol{D}_{\boldsymbol{k}_{m}}(\tau_{m})\dots\rangle\,,\label{cntra}
\end{align}
with two matrix vertices $\boldsymbol{s}_{\boldsymbol{k}}^{l}$ of
type $\boldsymbol{s}_{\boldsymbol{k}}^{x(y)}=\boldsymbol{U}_{\boldsymbol{k}}^{+}\boldsymbol{p}_{\boldsymbol{k}}^{x(y)}\boldsymbol{U}_{\boldsymbol{k}}^{\phantom{+}}$
(incoming) and one vertex $\boldsymbol{s}_{\boldsymbol{k}}^{\alpha}=\boldsymbol{U}_{\boldsymbol{k}}^{+}\boldsymbol{\mathfrak{F}}_{\boldsymbol{k}}^{\alpha}\boldsymbol{U}_{\boldsymbol{k}}^{\phantom{+}}$
(outgoing), which are the transforms to the diagonal boson representation
of the in-plane components of the polarization and of the force, respectively.

Symmetry restricts the phonon modes which the Raman force can act
on. Due to momentum conservation, the expression resulting from Eq.
(\ref{cntra}) involves a single $\boldsymbol{k}$-integration over
a product of a function of $\epsilon_{\boldsymbol{k}i}$, two $\boldsymbol{p}_{\boldsymbol{k}}^{x(y)}$,
and one $\boldsymbol{\mathfrak{F}}_{\boldsymbol{k}}^{\alpha}$. From
the previously listed symmetries, this integrand is (anti)symmetric
under $k_{x}\rightarrow-k_{x}$ for the (longitudinal)transverse mode.
I.e., the Raman force acts only on the transverse mode, marking it
as ``Raman active''. This allows Eq. (\ref{cntra}) to be considered
for $\alpha=0$ only. The superscript $\alpha$ on $\chi_{2}$ is
dropped hereafter. We mention that using a similar argument, only
the longitudinal mode is ``optically active''.

The time ordering allows for normal
\begin{equation}
-\langle T_{\tau}(D_{\boldsymbol{k}\mu}(\tau)D_{\boldsymbol{k}'\nu}^{\dagger})\rangle=\delta_{\mu\nu}\delta_{\boldsymbol{k}\boldsymbol{k}'}G_{\mu}(\boldsymbol{k},\tau)\,,\label{eq:5}
\end{equation}
as well as anomalous contractions
\begin{align}
-\langle T_{\tau}(D_{\boldsymbol{k}\mu}(\tau)D_{\boldsymbol{k}'\nu})\rangle= & -\langle T_{\tau}(D_{\boldsymbol{k}\mu}(\tau)D_{-\boldsymbol{k}'\bar{\nu}}^{\dagger})\rangle\nonumber \\
= & \,\delta_{\mu\bar{\nu}}\delta_{\boldsymbol{k},-\boldsymbol{k}'}G_{\mu}(\boldsymbol{k},\tau)\,,\label{eq:6}
\end{align}
where $\mu=1,..4$, and we define $\bar{\nu}=((\nu+1)$ mod $4)+1$,
which maps 1,2,3,4$\rightarrow$3,4,1,2. Finally, the adjoint anomalous
contractions satisfy $-\langle T_{\tau}(D_{\boldsymbol{k}\mu}^{\dagger}(\tau)\allowbreak D_{\boldsymbol{k}'\nu}^{\dagger})\rangle{=}\allowbreak-\langle T_{\tau}(D_{-\boldsymbol{k}\bar{\mu}}(\tau)\allowbreak D_{\boldsymbol{k}'\nu}^{\dagger})\rangle{=}\allowbreak\delta_{\mu\bar{\nu}}\delta_{\boldsymbol{k},-\boldsymbol{k}'}\allowbreak G_{\mu}(\boldsymbol{k},\tau)$,
using $\delta_{\bar{\mu}\nu}{=}\delta_{\mu\bar{\nu}}$.

The diagram for $\chi_{2}$ is a closed loop of three bilinear vertices
$\boldsymbol{s}_{\boldsymbol{k}}^{l}$, linked by a normal or anomalous
contraction. Since the anomalous Green's functions are related to
the normal ones by a mere shift of either the right or the left matrix
indices $\mu\rightarrow\bar{\mu}$ and a flip $\boldsymbol{k}'\rightarrow-\boldsymbol{k}'$
of the sign of the momentum, the necessary contractions can be reduced
to only normal ones, by relabeling the summation indices appropriately
and by using a symmetrized vertex $s_{\boldsymbol{k}\mu\nu}^{l}\rightarrow t_{\boldsymbol{k}\mu\nu}^{l}=(s_{\boldsymbol{k}\mu\nu}^{l}+s_{-\boldsymbol{k}\bar{\nu}\bar{\mu}}^{l})$.
With some algebra, one can show that $s_{-\boldsymbol{k}\bar{\nu}\bar{\mu}}^{l}=s_{\boldsymbol{k}\mu\nu}^{l}$,
which simplifies $t_{\boldsymbol{k}\mu\nu}^{l}=2s_{\boldsymbol{k}\mu\nu}^{l}$.
Summarizing, all diagrams can be generated using the matrix vertices
$t_{\boldsymbol{k}\mu\nu}^{l}$ and a 'single-arrowed' normal Green's
function $\boldsymbol{G}(\boldsymbol{k},i\omega_{n})$ of diagonal
matrix shape, with entries
\begin{align}
\boldsymbol{G}(\boldsymbol{k},i\omega_{n})= & \delta_{\mu\nu}G_{\mu}(\boldsymbol{k},i\omega_{n})\label{G0}\\
G_{\mu=1,2}(\boldsymbol{k},i\omega_{n})= & \frac{1}{i\omega_{n}-\epsilon_{\boldsymbol{k}i}}\nonumber \\
G_{\mu=3,4}(\boldsymbol{k},i\omega_{n})= & G_{\mu=1,2}(-\boldsymbol{k},-i\omega_{n})\,.\nonumber 
\end{align}
This is shown in Fig. (\ref{fig:RFdiag}). Both incoming frequencies
are kept as independent variables.

In principle, evaluation of Fig. (\ref{fig:RFdiag}) involves only
collecting three residues from the poles from the product of the Greens
functions. In practice, however, evaluating this ``brute-force''
leads to unmanageable results. This is due to the diagram summing
over $4^{3}=64$ flavors from the distribution of indices $\mu=1,..4$
for each Greens functions. Collecting their residues generates a sum
of $192$ rational functions of $\omega_{1,2}$. Simplification of
them remains unsatisfactory, owing in part to the mixed occurrence
of matrix elements from the $\boldsymbol{U}_{\boldsymbol{k}}^{+}\boldsymbol{p}_{\boldsymbol{k}}^{x(y)}\boldsymbol{U}_{\boldsymbol{k}}^{\phantom{+}}$
and $\boldsymbol{U}_{\boldsymbol{k}}^{+}\boldsymbol{\mathfrak{F}}_{\boldsymbol{k}}^{\alpha}\boldsymbol{U}_{\boldsymbol{k}}^{\phantom{+}}$
vertices. Therefore we present the result for $\chi_{2}^{l}(\omega_{1},\omega_{2})$
in a matrix notation
\begin{align}
\lefteqn{\chi_{2}^{l}(\omega_{1},\omega_{2})=\sum_{\boldsymbol{k},\mu}\mathrm{Tr}(}\label{chiRF}\\
 & \hphantom{a}\boldsymbol{G}(\boldsymbol{k},z_{1}{+}z_{2}{+}\varepsilon_{\boldsymbol{k}\mu})\,\boldsymbol{t}_{\boldsymbol{k}}^{l}\,\boldsymbol{G}(\boldsymbol{k},z_{1}{+}\varepsilon_{\boldsymbol{k}\mu})\,\boldsymbol{W}_{\mu}(\boldsymbol{t}_{\boldsymbol{k}}^{l},\boldsymbol{t}_{\boldsymbol{k}}^{0})+\nonumber \\
 & \hphantom{a}\boldsymbol{G}(\boldsymbol{k},z_{2}{+}\varepsilon_{\boldsymbol{k}\mu})\,\boldsymbol{W}_{\mu}(\boldsymbol{t}_{\boldsymbol{k}}^{l},\boldsymbol{t}_{\boldsymbol{k}}^{l})\,\boldsymbol{G}(\boldsymbol{k},-z_{1}{+}\varepsilon_{\boldsymbol{k}\mu})\,\boldsymbol{t}_{\boldsymbol{k}}^{0}+\nonumber \\
 & \hphantom{a}\boldsymbol{W}_{\mu}(\boldsymbol{t}_{\boldsymbol{k}}^{0},\boldsymbol{t}_{\boldsymbol{k}}^{l})\,\boldsymbol{G}(\boldsymbol{k},-z_{2}{+}\varepsilon_{\boldsymbol{k}\mu})\,\boldsymbol{t}_{\boldsymbol{k}}^{l}\,\boldsymbol{G}(\boldsymbol{k},-z_{1}{-}z_{2}{+}\varepsilon_{\boldsymbol{k}\mu}))\,,\hphantom{a}\nonumber 
\end{align}
with $z_{i}=\omega_{i}+i\,\eta$, $\eta>0$ and $l=x(y)$. Moreover
\begin{align}
\varepsilon_{\boldsymbol{k}\mu=1,..4} & =(\epsilon_{\boldsymbol{k}1},\epsilon_{\boldsymbol{k}2},-\epsilon_{\boldsymbol{k}1},-\epsilon_{\boldsymbol{k}2})\\
N_{\boldsymbol{k}\mu=1,..4} & =(n(\epsilon_{\boldsymbol{k}1}),n(\epsilon_{\boldsymbol{k}2}),1{+}n(\epsilon_{\boldsymbol{k}1}),1{+}n(\epsilon_{\boldsymbol{k}2}))\,,\label{eq:bosefact}
\end{align}
where $n(\epsilon)=1/(\exp(\epsilon/T)-1)$ is the Bose function.
The matrix elements of $\boldsymbol{W}$ for fixed $\mu$ are defined
as
\begin{equation}
[\boldsymbol{W}_{\mu}(\boldsymbol{X},\boldsymbol{Y})]_{\alpha\beta}=X_{\alpha\mu}N_{\boldsymbol{k}\mu}Y_{\mu\beta}\,.
\end{equation}
On a formal level, Eq. (\ref{chiRF}) clearly displays the structure
resulting from the integration over the internal frequency. On a practical
level, this expression is readily accessible to computational treatment.

At point we briefly mention that in principle one could speculate about the role of
more involved higher order interaction vertices which might also contribute to the
RF. E.g., the polarization vertices might display a magnetostrictive correction and
vice versa for the spin phonon coupling. This would imply additional diagrams with
additional unknown coupling parameters for the RF. At the level of the present
study, we refrain from incorporating such extensions \citep{LObub}.

\begin{figure*}[t]
\begin{centering}
\begin{minipage}[c]{0.34\textwidth}%
\begin{center}
  \includegraphics[width=1\textwidth]{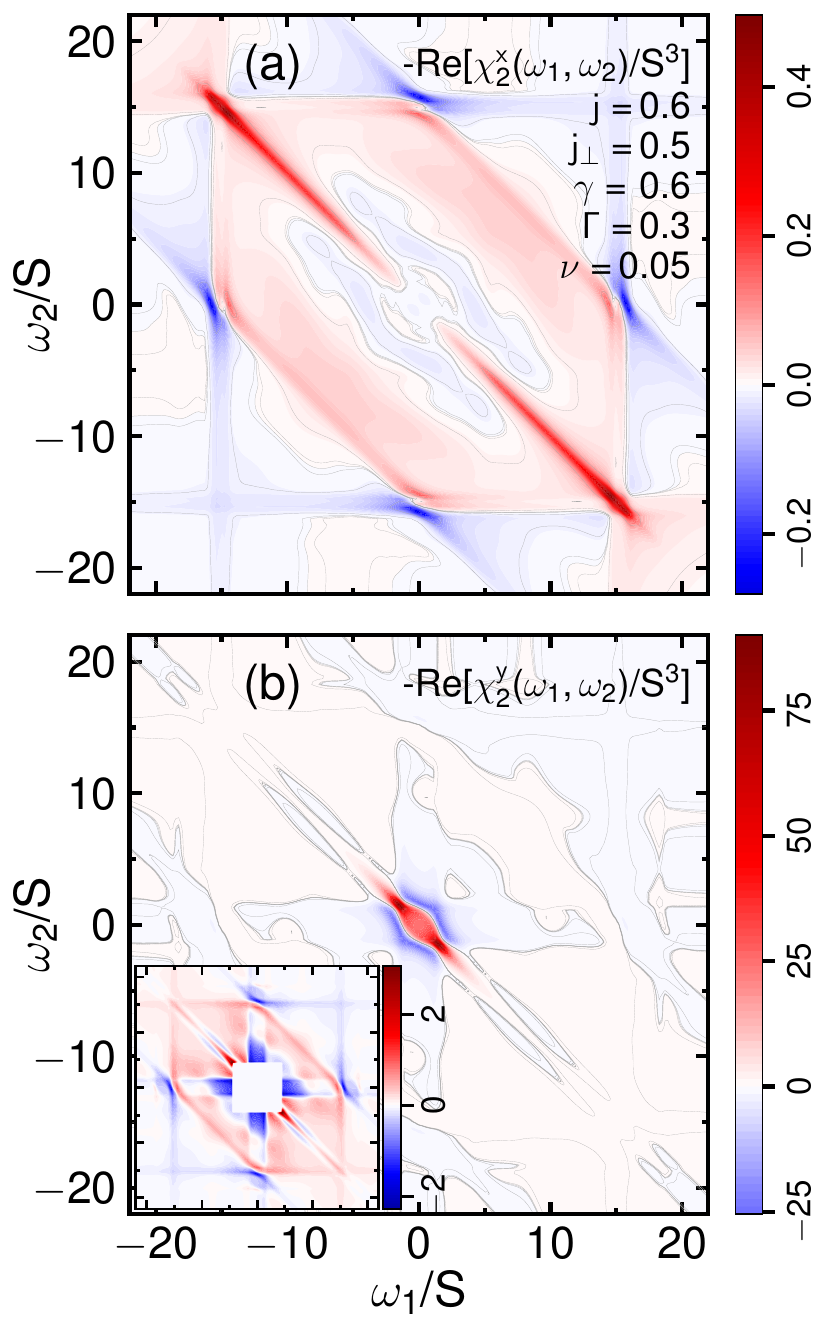}
\par\end{center}%
\end{minipage}%
\begin{minipage}[c]{0.66\textwidth}%
  \includegraphics[width=0.5\textwidth]{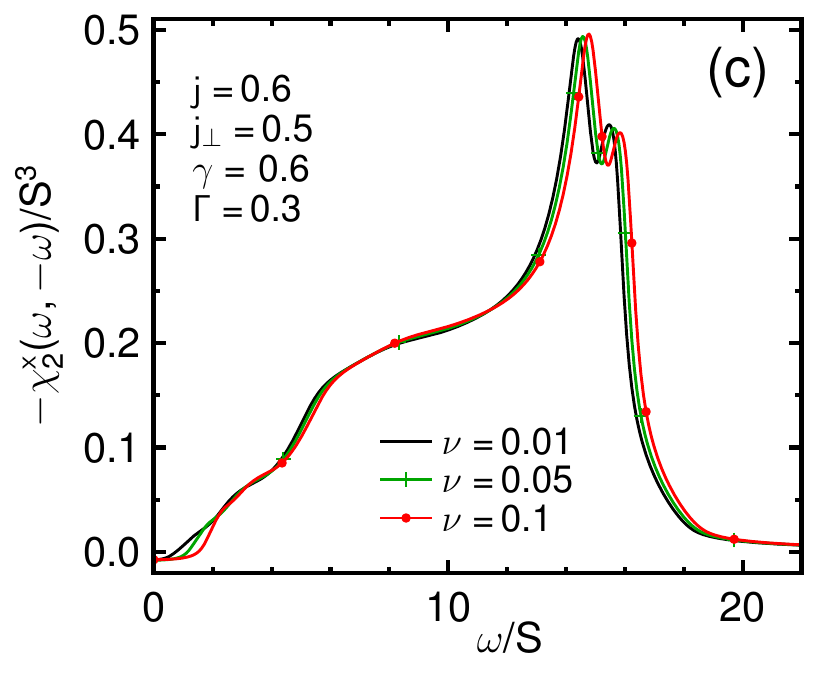}\includegraphics[width=0.5\textwidth]{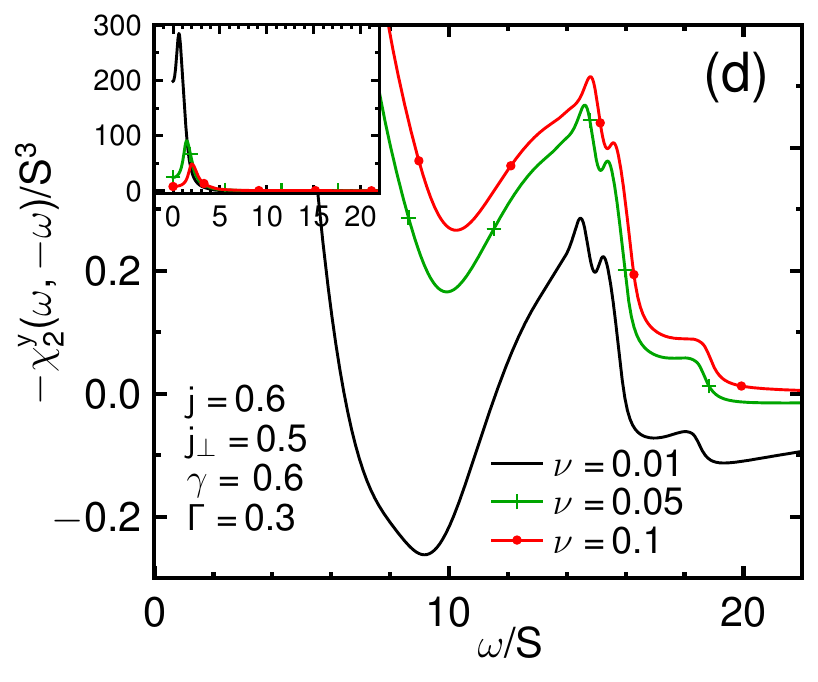}

  \includegraphics[width=0.5\textwidth]{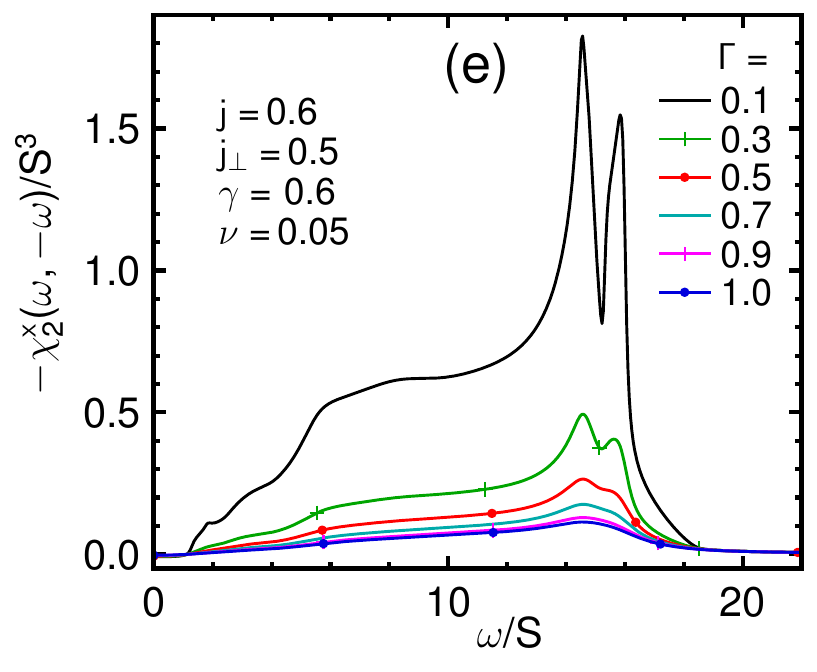}\includegraphics[width=0.5\textwidth]{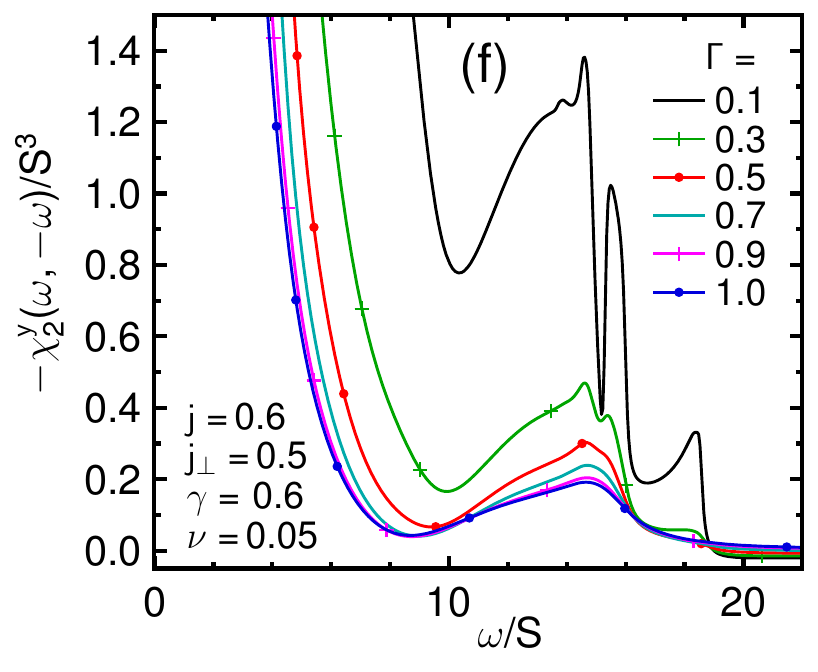}%
\end{minipage}
\par\end{centering}
\caption{Various views of the RF, all at fixed exchange parameters $j=0.6$,
$j_{\perp}=.5$, and $\gamma=0.6$. (a), (b): Contours of the real
part of the RF $\chi_{2}^{l}(\omega_{1},\omega_{2})$ in the 2D frequency
plane for $l=x[y]$ in (a){[}(b){]} at $\Gamma=0.3$ and $\nu=0.05$.
Contour lines added at $\pm(.1,.01,.001,0)$ for better visibility
of low amplitude structures. Inset in (b): Accentuation of lower intensities
contours on identical $\omega_{1,2}$-range, masking small-$\omega_{1,2}$
region. (c), (d): RF on the rectification line $\chi_{2}^{l}(\omega,-\omega)$
versus $\nu$ at $\Gamma=0.3$ for $l=x[y]$ in (c){[}(d){]}. Inset
in (d): Display of complete y-axis showing low-$\omega$ peak. (e),
(f): RF on the rectification line $\chi_{2}^{l}(\omega,-\omega)$ versus
$\Gamma$ at $\nu=0.05$ for $l=x[y]$ in (e){[}(f){]}. Low-$\omega$
peak in (f), similar to inset in (d), not displayed. \label{fig:RFcombo}}
\end{figure*}

\subsection*{Discussion of the Raman force}

Eq. (\ref{chiRF}) is a central result of this work. In Fig. \ref{fig:RFcombo}
we discuss several of its relevant features, using one fixed set of
exchange and magnetoelectric coupling parameters. First, we consider
$\mathrm{Re}[\chi_{2}^{l}(\omega_{1},\omega_{2})]$ in a two-dimensional
$\omega_{1,2}$-plane in Figs. \ref{fig:RFcombo}(a) for $l{=}x$
and \ref{fig:RFcombo}(b) for $l{=}y$. Obviously, $\mathrm{Re}[\chi_{2}^{l}(\omega_{1},\omega_{2})]$
is not only symmetric perpendicular to the diagonal $\omega_{1}{=}\omega_{2}$,
as required by intrinsic permutation symmetry, but also perpendicular
to the antidiagonal $\omega_{2}{=}-\omega_{1}$, the so-called rectification
line. As displayed for completeness in App. \ref{sec:A1}, the $\mathrm{Im}[\chi_{2}^{l}(\omega_{1},\omega_{2})]$
is also symmetric regarding $\omega_{1}\leftrightarrow\omega_{2}$,
but antisymmetric perpendicular to $\omega_{2}=\allowbreak-\omega_{1}$.
In turn, $\chi_{2}^{l}(\omega_{1},\omega_{2})$ is completely real
on the rectification line. From Eq. (\ref{chiRF}), the global $\omega_{1,2}$-scale
of $\chi_{2}^{l}(\omega_{1},\omega_{2})$ is set by $\sim2\mathrm{\,max}(\epsilon_{{\bf k}i})$.

Figs. \ref{fig:RFcombo}(a), (b) show that $\mathrm{Re}[\chi_{2}^{l}(\omega_{1},\omega_{2})]$
peaks when traversing the antidiagonal in a perpendicular direction.
Speaking more loosely, the rectification line is a dominant feature
in the 2D $\omega_{1,2}$-plane. This property has been analyzed in
great detail for response functions of 2D nonlinear optical spectroscopy
in Kitaev magnets \citep{Brenig2024}, spiral magnets \citep{Brenig2025},
XXZ chains \citep{Brenig2025b}, and also in the very different context
of shift and injection currents \citep{Parker2019,Sipe2000,Fei2020,Ishizuka2022,Raj2024}.
There it has been shown that in the vicinity of the rectification
line, three-point functions as in Fig. \ref{fig:RFdiag} can asymptotically
be expressed as the product of some finite, form-factor weighted two-particle
density of states multiplied by a Lorentzian $\sim i/(\omega_{\perp}+i\Gamma)$,
where $\omega_{\perp}$ is a frequency variable, perpendicular to
and zero on the rectification line. $\Gamma>0$ is a damping rate,
which for free particles reduces to the causal broadening $\eta$.
Therefore, $\chi_{2}^{l}(\omega,-\omega)$ scales roughly $\propto1/\text{\ensuremath{\Gamma}}$,
i.e., it is \emph{infinite} for free one-body excitations. To render
this finite, it has become customary to replace $\eta$ in Eq. (\ref{chiRF})
by a finite \emph{physical} damping rate $\Gamma$ \citep{Brenig2024,Brenig2025,Brenig2025b,Parker2019,Sipe2000,Fei2020,Ishizuka2022,Raj2024}.
Corrections beyond LSWT, e.g., magnon self-energies will provide for
$\Gamma$, but also scattering from other degrees of freedom. Here
we keep $\Gamma\ll\mathrm{max}(\epsilon_{{\bf k}i})$ as a free, momentum
and temperature independent parameter.

From the colorbar of Figs. \ref{fig:RFcombo}(a) and (b), it is evident
that $\chi_{2}^{l}(\omega_{1},\omega_{2})$ is strongly anisotropic,
with the scale of the response much larger for $l=y$ than for $l=x$.
Note that we have chosen a pitch $\boldsymbol{q}$ along the $x$-axis.
Therefore, the classical polarization $\boldsymbol{P}$ points along
$y$. Figs. \ref{fig:RFcombo}(b) also shows that the large intensity
in $\chi_{2}^{y}(\omega_{1},\omega_{2})$ is confined to low $\omega_{1,2}$,
implying that it results from energies in the vicinity of the zeros
of the magnon dispersion. I.e., this intensity is sensitive to the
coherence-length gap $\propto\sqrt{\nu}$. Masking the large values
at low-$\omega_{1,2}$ by a rectangle, as in the inset of Fig. \ref{fig:RFcombo}(b),
features of $\chi_{2}^{y}(\omega_{1},\omega_{2})$ surface which are
similar to those of Fig. \ref{fig:RFcombo}(a).

Next, in Figs. \ref{fig:RFcombo}(c)-(f) we consider the 1D slice
$\mathrm{Re}[\allowbreak\chi_{2}^{l}(\omega,\allowbreak-\omega)]$.
This is the fully rectified DC component of the RF. I.e., it is this
component which can induce a \emph{pure} DECP. In panels (c) and (d),
the dependence on the low energy cut-off $\nu$, Eq. (\ref{eq:hent1}),
is displayed for electric fields polarized into the $x$- and $y$-direction,
in (c) and (d), respectively. Clearly, for the $x$-direction, which
contains no relevant low-$\omega$ intensity a priori, the effect
of $\nu$ is weak and visible only through a small gap $\sim O(2\sqrt{\nu})$
at low energies and a negligible shift of the upper spectral edge.
For the $y$-polarization this is different. As noted already for
Fig. \ref{fig:RFcombo}(b), this direction is dominated by the Goldstone
zeros and therefore it varies significantly with $\nu$. To display
this, panel (d) is divided into a main figure and an inset, showing
the lower and the complete intensity range, respectively. Indeed,
there is a large peak in the RF at low-$\omega$, the height of which
can be cut off with increasing $\nu$. At larger $\omega$ there are
less intense structures, similar to those in the $x$-direction. For
both polarization directions, these structures exhibit van-Hove features,
smeared by $\Gamma$, and form factor weighted by the polarization
and spin-phonon matrix elements. The upper two-magnon band edge at
$2\mathrm{\,max}(\epsilon_{{\bf k}i})$ clearly shows up for both,
$x$- and $y$-polarization. It is tempting to speculate that in an
experimental situation, the strong anisotropy may be observable, potentially
with some six-fold angular variation if magnetic domain formation
exists.

In Figs. \ref{fig:RFcombo}(e) and (f) we display the variation of
the RF with the magnon damping rate $\Gamma$. \emph{Cum granu salis}
the two panel show exactly the scaling of the RF with $1/\Gamma$,
which has been anticipated already in our discussion of $\chi_{2}^{l}(\omega_{1},\omega_{2})$
in the 2D frequency plane in panels (a) and (b). This behavior is
another remarkable feature of the RF. From an experimental point of
view, it suggests that sample quality may be a decisive factor, determining
the strength of the RF. Regarding theory, it implies that dressing
the diagram of Fig. \ref{fig:RFdiag} with renormalizations beyond
LSWT or extrinsic interactions may be an interesting future direction
to investigate. This is beyond the present study. Since no additional
information can be obtained from it, we have not included an inset
for the low-$\omega$ peak in panel (f), similar to that in (d).

Finally, we mention without any explicit analysis, that because of
the Bose factors in Eq. (\ref{eq:bosefact}), increasing the temperature
in Figs. \ref{fig:RFcombo}(a)-(f) has the usual effect of increasing
the weight of intensities at lower frequencies.

\section{Conclusion\label{sec:Conclusion}}

It is well established that by nonlinear coupling of electric laser
fields to the electronic degrees of freedom of many solids, so called
Raman forces can be generated which allow to excite coherent phonon
modes, observable in pump-probe spectroscopy. Our work shows that
this concept can be extended into a completely different area, namely
that of the interaction of electric fields with the spin degrees of
freedom of multiferroic quantum magnets. We have explicitly detailed
this for a simplified AB-stacked bilayer spin-model resembling the
emergent class of layered vdW TMDs which are type-II multiferroics.
For suitable exchange matrix elements, the bilayer is in an incommensurate
spiral state which has a finite electric polarization and allows for
coupling of magnons to the laser field via the spin-current interaction.
For the phonons, we have focused on the shear modes of the bilayer
which couple magnetoelastically to the quantum magnet. From this setup,
we have derived a Raman force that is highly anisotropic with respect
to the direction of the spiral pitch vector, scales proportional to
the lifetime of the magnons, and has relevant spectral structure up
to the maximum of the two-magnon energy. Due to the current status
of \emph{ab-initio} approaches to the vdW TMDs, our analysis clearly
remains a model study. However, we believe that its main ideas will
qualitatively withstand a more refined microscopic modeling of these
materials. From an experimental perspective, it could be interesting
to consider pump-probe spectrocopy on TMD bilayers and potential Moir\'e
realizations thereof, in particular, traversing multiferroic transitions,
in order to compare the response in and out of the ordered phase.
\begin{acknowledgments}
This research was supported in part by the DFG through Project A02
of SFB 1143 (project-id 247310070). I acknowledge kind hospitality
of the PSM, Dresden.
\end{acknowledgments}

\appendix

\section{Imaginary part of the RF\label{sec:A1}}

\begin{figure}[!t]
\centering{}\vspace{0mm}
\includegraphics[width=0.57\columnwidth]{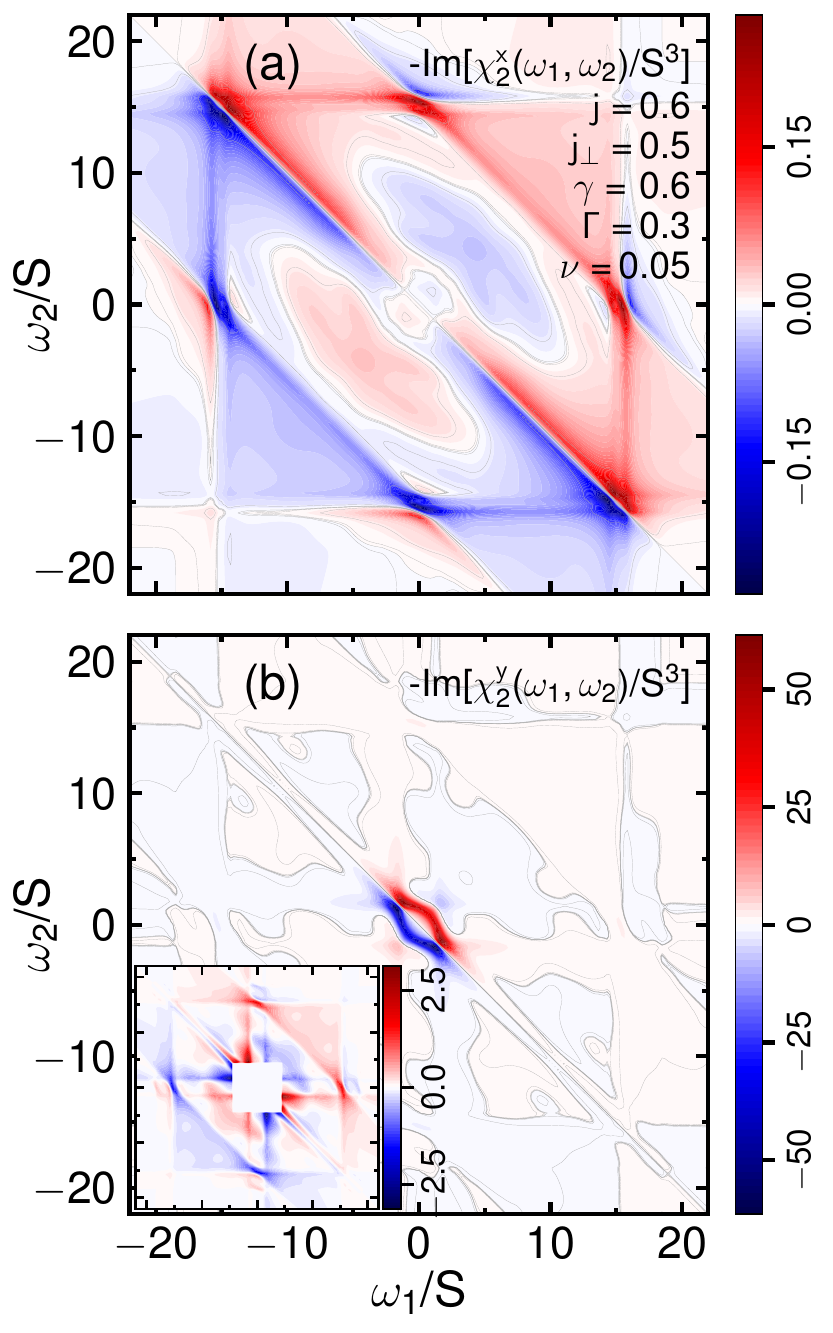}\vspace{-4mm}
\caption{(a), (b): Contours of the imaginary part of the RF $\chi_{2}^{l}(\omega_{1},\omega_{2})$
in the 2D frequency plane for $l=x[y]$ in (a){[}(b){]} at $j=0.6$,
$j_{\perp}=.5$, $\gamma=0.6$, $\Gamma=0.3$, and $\nu=0.05$. Contour
lines added at $\pm(.1,.01,.001,0)$ for better visibility of low
amplitude structures. Inset in (b): Accentuation of lower intensities
contours on identical $\omega_{1,2}$-range, masking small-$\omega_{1,2}$
region.\label{fig:2DImRF}\vspace{-6mm}
}
\end{figure}

This appendix is intended solely to display additional data of the
RF, namely the imaginary part of $\chi_{2}^{l}(\omega_{1},\omega_{2})$
in a two-dimensional $\omega_{1,2}$-plane in Figs. \ref{fig:2DImRF}(a)
for $l{=}x$ and \ref{fig:2DImRF}(b) for $l{=}y$. The main point
is that in contrast to the real part, the imaginary part is antisymmetric
perpendicular to\vspace{-4mm}
 the rectification line.

%\cleardoublepage{}


\begin{thebibliography}{0}%
\makeatletter
\providecommand \@ifxundefined [1]{%
 \@ifx{#1\undefined}
}%
\providecommand \@ifnum [1]{%
 \ifnum #1\expandafter \@firstoftwo
 \else \expandafter \@secondoftwo
 \fi
}%
\providecommand \@ifx [1]{%
 \ifx #1\expandafter \@firstoftwo
 \else \expandafter \@secondoftwo
 \fi
}%
\providecommand \natexlab [1]{#1}%
\providecommand \enquote  [1]{``#1''}%
\providecommand \bibnamefont  [1]{#1}%
\providecommand \bibfnamefont [1]{#1}%
\providecommand \citenamefont [1]{#1}%
\providecommand \href@noop [0]{\@secondoftwo}%
\providecommand \href [0]{\begingroup \@sanitize@url \@href}%
\providecommand \@href[1]{\@@startlink{#1}\@@href}%
\providecommand \@@href[1]{\endgroup#1\@@endlink}%
\providecommand \@sanitize@url [0]{\catcode `\\12\catcode `\$12\catcode
  `\&12\catcode `\#12\catcode `\^12\catcode `\_12\catcode `\%12\relax}%
\providecommand \@@startlink[1]{}%
\providecommand \@@endlink[0]{}%
\providecommand \url  [0]{\begingroup\@sanitize@url \@url }%
\providecommand \@url [1]{\endgroup\@href {#1}{\urlprefix }}%
\providecommand \urlprefix  [0]{URL }%
\providecommand \Eprint [0]{\href }%
\providecommand \doibase [0]{https://doi.org/}%
\providecommand \selectlanguage [0]{\@gobble}%
\providecommand \bibinfo  [0]{\@secondoftwo}%
\providecommand \bibfield  [0]{\@secondoftwo}%
\providecommand \translation [1]{[#1]}%
\providecommand \BibitemOpen [0]{}%
\providecommand \bibitemStop [0]{}%
\providecommand \bibitemNoStop [0]{.\EOS\space}%
\providecommand \EOS [0]{\spacefactor3000\relax}%
\providecommand \BibitemShut  [1]{\csname bibitem#1\endcsname}%
\let\auto@bib@innerbib\@empty
%</preamble>
\end{thebibliography}%


\begin{thebibliography}{99}
\bibitem{Weiner2009}A. M. Weiner, Ultrafast Optics, Wiley, Series
in Pure and Applied Optics, First Edition, 2009.

\bibitem{Cavalleri2001}A. Cavalleri, Cs. T\'oth, C. W. Siders, J.
A. Squier, F. R\'aksi, P. Forget, and J. C. Kieffer, Femtosecond
Structural Dynamics in VO$_{2}$ during an Ultrafast Solid-Solid Phase
Transition, Phys. Rev. Lett. \textbf{87}, 237401 (2001).

\bibitem{Chollet2005}M. Chollet et al., Gigantic Photoresponse in
$\nicefrac{1}{4}$-Filled-Band Organic Salt (EDO-TTF)2 PF6, Science
\textbf{307}, 86 (2005).

\bibitem{Fushitani2008}M. Fushitani, Applications of pump-probe spectroscopy,
Annu. Rep. Prog. Chem., Sect. C: Phys. Chem. \textbf{104}, 272 (2008).

\bibitem{Saiki2017}T. Saiki, N. Yamaguchi, Y. Obata, Y. Kakesu, T.
Kajita, and T. Katsufuji, Photoinduced phase transitions over three
phases in Ba$_{0.9}$Sr$_{0.1}$V$_{13}$O$_{18}$, Phys. Rev. B \textbf{96},
035133 (2017).

\bibitem{Katsufuji2023}T. Katsufuji, T. Yoshida, K. Akimoto, S. Okubo,
and T. Saiki, Pump-probe optical spectroscopy of hexagonal YMnO$_{3}$
over wide time and energy ranges, Phys. Rev. B \textbf{108}, 125131
(2023).

\bibitem{Cheng1991}T. K. Cheng, J. Vidal, H. J. Zeiger, G. Dresselhaus,
M. S. Dresselhaus, and E. P. Ippen, Mechanism for displacive excitation
of coherent phonons in Sb, Bi, Te, and Ti$_{2}$O$_{3}$, Applied
Physics Letters \textbf{59}, 1923 (1991).

\bibitem{Liu1995}Y. Liu, A. Frenkel, G. A. Garrett, J. F. Whitaker,
S. Fahy, C. Uher, and R. Merlin, Impulsive Light Scattering by Coherent
Phonons in LaAl O$_{3}$\,: Disorder and Boundary Effects, Phys. Rev.
Lett. \textbf{75}, 334 (1995).

\bibitem{Cho1990}G. C. Cho, W. K\"utt, and H. Kurz, Subpicosecond
time-resolved coherent-phonon oscillations in GaAs, Phys. Rev. Lett.
\textbf{65}, 764 (1990).

\bibitem{Pfeifer1992}T. Pfeifer, W. K\"utt, H. Kurz, and R. Scholz,
Generation and detection of coherent optical phonons in germanium,
Phys. Rev. Lett. \textbf{69}, 3248 (1992).

\bibitem{Yamamoto1994}A. Yamamoto, T. Mishina, Y. Masumoto, and M.
Nakayama, Coherent Oscillation of Zone-Folded Phonon Modes in GaAs-AlAs
Superlattices, Phys. Rev. Lett. \textbf{73}, 740 (1994).

\bibitem{Dekorsy1995}T. Dekorsy, H. Auer, C. Waschke, H. J. Bakker,
H. G. Roskos, H. Kurz, V. Wagner, and P. Grosse, Emission of Submillimeter
Electromagnetic Waves by Coherent Phonons, Phys. Rev. Lett. \textbf{74},
738 (1995).

\bibitem{Kuett1992}W. A. K\"utt, W. Albrecht, and H. Kurz, IEEE
J. Quantum Electron. \textbf{28}, 2434 (1992).

\bibitem{Merlin1997}R. Merlin, Generating coherent THz phonons with
light pulses, Solid State Communications \textbf{102}, 207 (1997).

\bibitem{Chwalek1990}J. M. Chwalek, C. Uher, J. F. Whitaker, G. A.
Mourou, J. Agostinelli, and M. Lelental, Femtosecond optical absorption
studies of nonequilibrium electronic processes in high-T$_{c}$ superconductors,
Applied Physics Letters \textbf{57}, 1696 (1990).

\bibitem{Mazin1994}I. I. Mazin, A. I. Liechtenstein, O. Jepsen, O.
K. Andersen, and C. O. Rodriguez, Displacive excitation of coherent
phonons in YBa$_{2}$Cu$_{3}$O$_{7}$, Phys. Rev. B \textbf{49},
9210 (1994).

\bibitem{Misochko2004}O. V. Misochko, M. V. Lebedev, N. Georgiev,
and T. Dekorsy, Coherent phonons in NdBa$_{2}$Cu$_{3}$O$_{7-x}$
single crystals: Optical-response anisotropy and hysteretic behavior,
J. Exp. Theor. Phys. \textbf{98}, 341 (2004).

\bibitem{Zeiger1992}H. J. Zeiger, J. Vidal, T. K. Cheng, E. P. Ippen,
G. Dresselhaus, and M. S. Dresselhaus, Theory for displacive excitation
of coherent phonons, Phys. Rev. B \textbf{45}, 768 (1992).

\bibitem{Scholz1993}R. Scholz, T. Pfeifer, and H. Kurz, Density-matrix
theory of coherent phonon oscillations in germanium, Phys. Rev. B
\textbf{47}, 16229 (1993).

\bibitem{Kuznetsov1994}A. V. Kuznetsov and C. J. Stanton, Theory
of Coherent Phonon Oscillations in Semiconductors, Phys. Rev. Lett.
\textbf{73}, 3243 (1994).

\bibitem{Garrett1996}G. A. Garrett, T. F. Albrecht, J. F. Whitaker,
and R. Merlin, Coherent THz Phonons Driven by Light Pulses and the
Sb Problem: What is the Mechanism?, Phys. Rev. Lett. \textbf{77},
3661 (1996).

\bibitem{Stevens2002}T. E. Stevens, J. Kuhl, and R. Merlin, Coherent
phonon generation and the two stimulated Raman tensors, Phys. Rev.
B \textbf{65}, 144304 (2002).

\bibitem{Ho2006}J. H. Ho, C. L. Lu, C. C. Hwang, C. P. Chang, and
M. F. Lin, Coulomb excitations in AA- and AB-stacked bilayer graphites,
Phys. Rev. B \textbf{74}, 085406 (2006).

\bibitem{Koshino2009}M. Koshino and E. McCann, Gate-induced interlayer
asymmetry in ABA-stacked trilayer graphene, Phys. Rev. B \textbf{79},
125443 (2009).

\bibitem{Zhang2019}M. Y. Zhang, Z. X. Wang, Y. N. Li, L. Y. Shi,
D. Wu, T. Lin, S. J. Zhang, Y. Q. Liu, Q. M. Liu, J. Wang, T. Dong,
and N. L. Wang, Light-Induced Subpicosecond Lattice Symmetry Switch
in MoTe$_{2}$, Phys. Rev. X \textbf{9}, 021036 (2019).

\bibitem{Fukuda2020}T. Fukuda, K. Makino, Y. Saito, P. Fons, A. V.
Kolobov, K. Ueno, and M. Hase, Ultrafast dynamics of the low frequency
shear phonon in 1T\textasciiacute -MoTe2, Applied Physics Letters
116, 093103 (2020).

\bibitem{Ji2021}S. Ji, O. Gr\r{a}n\"as, and J. Weissenrieder, Manipulation
of Stacking Order in Td -WTe$_{2}$ by Ultrafast Optical Excitation,
ACS Nano \textbf{15}, 8826 (2021).

\bibitem{Rostami2022}H. Rostami, Theory for shear displacement by
light-induced Raman force in bilayer graphene, Phys. Rev. B \textbf{106},
155405 (2022).

\bibitem{Fiebig2005}M. Fiebig, Revival of the magnetoelectric effect,
J. Phys. D: Appl. Phys. \textbf{38}, R123 (2005).

\bibitem{Tokura2014}Y. Tokura, S. Seki, and N. Nagaosa, Multiferroics
of spin origin, Rep. Prog. Phys. \textbf{77}, 076501 (2014).

\bibitem{Dong2015}S. Dong, J.-M. Liu, S.-W. Cheong, and Z. Ren, Multiferroic
materials and magnetoelectric physics: symmetry, entanglement, excitation,
and topology, Advances in Physics \textbf{64}, 519 (2015).

\bibitem{Dong2019}S. Dong, H. Xiang, and E. Dagotto, Magnetoelectricity
in multiferroics: a theoretical perspective, National Science Review
\textbf{6}, 629 (2019).

\bibitem{Botana2019}A. S. Botana and M. R. Norman, Electronic structure
and magnetism of transition metal dihalides: Bulk to monolayer, Phys.
Rev. Materials \textbf{3}, 044001 (2019).

\bibitem{Zhang2020}J. S. Zhang, Y. Xie, X. Q. Liu, A. Razpopov, V.
Borisov, C. Wang, J. P. Sun, Y. Cui, J. C. Wang, X. Ren, H. Deng,
X. Yin, Y. Ding, Y. Li, J. G. Cheng, J. Feng, R. Valent\'i, B. Normand,
and W. Yu, Giant pressure-enhancement of multiferroicity in CuBr$_{2}$,
Phys. Rev. Research \textbf{2}, 013144 (2020).

\bibitem{Amoroso2020}D. Amoroso, P. Barone, and S. Picozzi, Spontaneous
skyrmionic lattice from anisotropic symmetric exchange in a Ni-halide
monolayer, Nat Commun \textbf{11}, 5784 (2020).

\bibitem{Liu2020}H. Liu, X. Wang, J. Wu, Y. Chen, J. Wan, R. Wen,
J. Yang, Y. Liu, Z. Song, and L. Xie, Vapor Deposition of Magnetic
Van der Waals NiI$_{2}$ Crystals, ACS Nano \textbf{14}, 10544 (2020).

\bibitem{Ju2021}H. Ju, Y. Lee, K.-T. Kim, I. H. Choi, C. J. Roh,
S. Son, P. Park, J. H. Kim, T. S. Jung, J. H. Kim, K. H. Kim, J.-G.
Park, and J. S. Lee, Possible Persistence of Multiferroic Order down
to Bilayer Limit of van der Waals Material NiI$_{2}$, Nano Lett.
\textbf{21}, 5126 (2021).

\bibitem{Bikal2021}D. Bikaljevi\'c, C. Gonz\'alez-Orellana, M.
Pe\~na-D\'iaz, D. Steiner, J. Dreiser, P. Gargiani, M. Foerster,
M. \'A. Ni\~no, L. Aballe, S. Ruiz-Gomez, N. Friedrich, J. Hieulle,
L. Jingcheng, M. Ilyn, C. Rogero, and J. I. Pascual, Noncollinear
Magnetic Order in Two-Dimensional NiBr$_{2}$ Films Grown on Au(111),
ACS Nano \textbf{15}, 14985 (2021).

\bibitem{Song2022}Q. Song, C. A. Occhialini, E. Erge\c{c}en, B.
Ilyas, D. Amoroso, P. Barone, J. Kapeghian, K. Watanabe, T. Taniguchi,
A. S. Botana, S. Picozzi, N. Gedik, and R. Comin, Evidence for a single-layer
van der Waals multiferroic, Nature \textbf{602}, 601 (2022).

\bibitem{Sodequist2023}J. S{\o}dequist and T. Olsen, Type II multiferroic
order in two-dimensional transition metal halides from first principles
spin-spiral calculations, 2D Mater. \textbf{10}, 035016 (2023).

\bibitem{Lebedev2023}D. Lebedev, J. T. Gish, E. S. Garvey, T. K.
Stanev, J. Choi, L. Georgopoulos, T. W. Song, H. Y. Park, K. Watanabe,
T. Taniguchi, N. P. Stern, V. K. Sangwan, and M. C. Hersam, Electrical
Interrogation of Thickness-Dependent Multiferroic Phase Transitions
in the 2D Antiferromagnetic Semiconductor NiI$_{2}$, Adv Funct Materials
\textbf{33}, 2212568 (2023).

\bibitem{Jiang2023}Y. Jiang, Y. Wu, J. Zhang, J. Wei, B. Peng, and
C.-W. Qiu, Dilemma in optical identification of single-layer multiferroics,
Nature \textbf{619}, E40 (2023).

\bibitem{Gao2024}F. Y. Gao, X. Peng, X. Cheng, E. Vi\~nas Bostr\"om,
D. S. Kim, R. K. Jain, D. Vishnu, K. Raju, R. Sankar, S.-F. Lee, M.
A. Sentef, T. Kurumaji, X. Li, P. Tang, A. Rubio, and E. Baldini,
Giant chiral magnetoelectric oscillations in a van der Waals multiferroic,
Nature \textbf{632}, 273 (2024).

\bibitem{Song2025}Q. Song, S. Stavri\'c, P. Barone, A. Droghetti,
D. S. Antonenko, J. W. F. Venderbos, C. A. Occhialini, B. Ilyas, E.
Erge\c{c}en, N. Gedik, S.-W. Cheong, R. M. Fernandes, S. Picozzi,
and R. Comin, Electrical switching of a p-wave magnet, Nature \textbf{642},
64 (2025).

\bibitem{NLiu2024}N. Liu, C. Wang, C. Yan, C. Xu, J. Hu, Y. Zhang,
and W. Ji, Competing multiferroic phases in monolayer and few-layer
NiI$_{2}$, Phys. Rev. B \textbf{109}, 195422 (2024).

\bibitem{Bellaiche2023}X. Li, C. Xu, B. Liu, X. Li, L. Bellaiche,
and H. Xiang, Realistic Spin Model for Multiferroic NiI$_{2}$, Phys.
Rev. Lett. \textbf{131}, 036701 (2023).

\bibitem{Bennett2024}D. Bennett, G. Mart\'inez-Carracedo, X. He, J.
Ferrer, P. Ghosez, R. Comin, and E. Kaxiras, Stacking-Engineered Ferroelectricity
and Multiferroic Order in van der Waals Magnets, Phys. Rev. Lett.
\textbf{133}, 246703 (2024).

\bibitem{Antao2024}T. V. C. Ant\~ao, J. L. Lado, and A. O. Fumega,
Electric Field Control Of Moir\'e Skyrmion Phases in Twisted Multiferroic
NiI$_{2}$ Bilayers, Nano Lett. \textbf{24}, 15767 (2024).

\bibitem{Amini2024}M. Amini, A. O. Fumega, H. Gonz\'alez-Herrero,
V. Va\~no, S. Kezilebieke, J. L. Lado, and P. Liljeroth, Atomic-Scale
Visualization of Multiferroicity in Monolayer NiI$_{2}$, Advanced
Materials \textbf{36}, 2311342 (2024).

\bibitem{Katsura2005}H. Katsura, N. Nagaosa, and A. V. Balatsky,
Spin Current and Magnetoelectric Effect in Noncollinear Magnets, Phys.
Rev. Lett. \textbf{95}, 057205 (2005).

\bibitem{Wu2023}S. Wu, X. Chen, C. Hong, X. Hou, Z. Wang, Z. Sheng,
Z. Sun, Y. Guo, and S. Wu, Layer Thickness Crossover of Type-II Multiferroic
Magnetism in NiI$_{2}$, arXiv:2307.10686.

\bibitem{Wu2024}Y. Wu, Z. Zeng, H. Lu, X. Han, C. Yang, N. Liu, X.
Zhao, L. Qiao, W. Ji, R. Che, L. Deng, P. Yan, and B. Peng, Coexistence
of ferroelectricity and antiferroelectricity in 2D van der Waals multiferroic,
Nat Commun \textbf{15}, 8616 (2024).

\bibitem{Bog1lay}We note that for the case of single layer an analytic
form of the Bogoliubov transformation can be obtained.

\bibitem{Colpa1978}J. H. P. Colpa, Diagonalization of the quadratic
boson hamiltonian, Physica A: Statistical Mechanics and Its Applications
\textbf{93}, 327 (1978).

\bibitem{HH}B. I. Halperin and P. C. Hohenberg, Generalization of
Scaling Laws to Dynamical Properties of a System Near its Critical
Point, Phys. Rev. Lett. \textbf{19}, 700 (1967); Scaling Laws for
Dynamic Critical Phenomena, Phys. Rev. \textbf{177}, 952 (1969).

\bibitem{CHN}S. Chakravarty, B. I. Halperin, and D. R. Nelson, Two-dimensional
quantum Heisenberg antiferromagnet at low temperatures, Phys. Rev.
B \textbf{39}, 2344 (1989).

\bibitem{Grempel}D. R. Grempel, Comment on \textquotedbl Low-Temperature
Behavior of Two-Dimensional Quantum Antiferromagnets\textquotedbl ,
Phys. Rev. Lett. \textbf{61}, 1041 (1988).

\bibitem{Tyc}S. T\v{y}c, B. I. Halperin, and S. Chakravarty, Dynamic
Properties of a Two-Dimensional Heisenberg Antiferromagnet at Low
Temperatures, Phys. Rev. Lett. \textbf{62}, 835 (1989).

\bibitem{Makivic}M. Makivi\v{c} and M. Jarrell, Low-temperature
dynamics of the 2D spin-1/2 Heisenberg antiferromagnet: A quantum
Monte Carlo study, Phys. Rev. Lett. \textbf{68}, 1770 (1992).

\bibitem{Hasenfratz00}Hasenfratz, The correlation length of the Heisenberg
antiferromagnet with arbitrary spin S, Eur. Phys. J. B \textbf{13},
11 (2000).

\bibitem{Greven}O. P. Vajk, P. K. Mang, M. Greven, P. M. Gehring,
and J. W. Lynn, Quantum Impurities in the Two-Dimensional Spin One-Half
Heisenberg Antiferromagnet, Science \textbf{295}, 1691 (2002).

\bibitem{Kastner98}M. A. Kastner, R. J. Birgeneau, G. Shirane, and
Y. Endoh, Magnetic, transport, and optical properties of monolayer
copper oxides, Rev. Mod. Phys. \textbf{70}, 897 (1998).

\bibitem{Ronnow}H. M. R{\o}nnow, D. F. McMorrow, R. Coldea, A. Harrison,
I. D. Youngson, T. G. Perring, G. Aeppli, O. Sylju{\aa}sen, K. Lefmann,
and C. Rischel, Spin Dynamics of the 2D Spin 1/2 Quantum Antiferromagnet
Copper Deuteroformate Tetradeuterate (CFTD), Phys. Rev. Lett. \textbf{87},
037202 (2001).

\bibitem{Goff}A. M. Toader, J. P. Goff, M. Roger, N. Shannon, J.
R. Stewart, and M. Enderle, Spin Correlations in the Paramagnetic
Phase and Ring Exchange in La$_{2}$CuO$_{4}$, Phys. Rev. Lett. \textbf{94},
197202 (2005).

\bibitem{Takahashi}M. Takahashi, Modified spin-wave theory of a square-lattice
antiferromagnet, Phys. Rev. B \textbf{40}, 2494 (1989).

\bibitem{Auerbach1988}A. Auerbach and D. P. Arovas, Spin Dynamics
in the Square-Lattice Antiferromagnet, Phys. Rev. Lett. \textbf{61},
617 (1988).

\bibitem{Barabanov1992}A. F. Barabanov, and O. A. Starykh, Spherical
Symmetric Spin Wave Theory of Heisenberg Model, J. Phys. Soc. Jpn.
\textbf{61}, 704 (1992).

\bibitem{Woods2000}L. M. Woods and G. D. Mahan, Electron-phonon effects
in graphene and armchair (10,10) single-wall carbon nanotubes, Phys.
Rev. B \textbf{61}, 10651 (2000).

\bibitem{Butcher1990}P. N. Butcher, D. Cotter, The Elements of Nonlinear
Optics, Cambridge University Press, Cambridge, 1990.

\bibitem{Evans1966}W. A. B. Evans, A general analytic continuation
response formalism for large quantum systems, Proc. Phys. Soc. \textbf{88},
723 (1966).

\bibitem{Rostami2021}H. Rostami, M. I. Katsnelson, G. Vignale, and
M. Polini, Gauge invariance and Ward identities in nonlinear response
theory, Annals of Physics \textbf{431}, 168523 (2021).

\bibitem{LObub} Higher order interaction vertices imply less than
  three-point correlation functions which are subdominant on the
  rectification line.

\bibitem{Brenig2024}W. Brenig and O. Krupnitska, Response functions
for electric field induced two-dimensional nonlinear spectroscopy
in a Kitaev magnet, J. Phys.: Condens. Matter \textbf{36}, 505806
(2024).

\bibitem{Brenig2025}W. Brenig, Two-dimensional nonlinear optical
response of a spiral magnet, Phys. Rev. B \textbf{112}, 115132 (2025).

\bibitem{Brenig2025b}W. Brenig, Two-Dimensional Nonlinear Dynamical
Response of the Magnetoelectrically Driven Dimerized Spin-$1{/}2$
Chain, arXiv:2507.17823.

\bibitem{Parker2019}D. E. Parker, T. Morimoto, J. Orenstein, and
J. E. Moore, Diagrammatic approach to nonlinear optical response with
application to Weyl semimetals, Phys. Rev. B \textbf{99}, 045121 (2019).

\bibitem{Sipe2000}J. E. Sipe and A. I. Shkrebtii, Second-order optical
response in semiconductors, Phys. Rev. B \textbf{61}, 5337 (2000).

\bibitem{Fei2020}R. Fei, W. Song, and L. Yang, Giant photogalvanic
effect and second-harmonic generation in magnetic axion insulators,
Phys. Rev. B \textbf{102}, 035440 (2020).

\bibitem{Ishizuka2022}H. Ishizuka and M. Sato, Large Photogalvanic
Spin Current by Magnetic Resonance in Bilayer Cr Trihalides, Phys.
Rev. Lett. \textbf{129}, 107201 (2022).

\bibitem{Raj2024}A. Raj, S. Chaudhary, and G. A. Fiete, Photogalvanic
response in multi-Weyl semimetals, Phys. Rev. Research \textbf{6},
013048 (2024).

\end{thebibliography}
\end{document}